\documentclass[twocolumn]{aastex631}

\usepackage{hyperref}
\usepackage{amsmath,amstext}
\usepackage[figure,figure*]{hypcap}
\usepackage{amssymb}
\usepackage{nicefrac}
\usepackage{graphicx}

\usepackage{wrapfig}

\shorttitle{Massive Black Hole Merger Host Galaxy Stellar Kinematics}
\begin{document}

\title{Signatures of Massive Black Hole Merger Host Galaxies from Cosmological Simulations II: \\ Unique Stellar Kinematics in Integral Field Unit Spectroscopy}

\correspondingauthor{Jaeden Bardati}
\email{jbardati@caltech.edu}

\author[0009-0002-8417-4480]{Jaeden Bardati}
\affiliation{Department of Physics \& Astronomy, Bishop's University, Sherbrooke, QC J1M 1Z7, Canada}
\affiliation{The Division of Physics, Mathematics and Astronomy, California Institute of Technology, Pasadena, CA 91125, USA}

\author[0000-0001-8665-5523]{John J. Ruan}
\affiliation{Department of Physics \& Astronomy, Bishop's University, Sherbrooke, QC J1M 1Z7, Canada}

\author[0000-0001-6803-2138]{Daryl Haggard}
\affiliation{McGill Space Institute and Department of Physics, McGill University, 3600 rue University, Montreal, Quebec, H3A 2T8, Canada}

\author[0000-0002-4353-0306]{Michael Tremmel}
\affiliation{School of Physics, University College Cork, Cork, Ireland}

\author[0009-0007-3541-435X]{Patrick Horlaville}
\affiliation{Department of Physics \& Astronomy, Bishop's University, Sherbrooke, QC J1M 1Z7, Canada}

\begin{abstract}
Secure methods for identifying the host galaxies of individual massive black hole (MBH) binaries and mergers detected by gravitational wave experiments such as LISA and Pulsar Timing Arrays are currently lacking, but will be critical to a variety of science goals. Recently in \citet[][Paper I]{Bardati_2024}, we used the Romulus25 cosmological simulation to show that MBH merger host galaxies have unique morphologies in imaging, due to their stronger bulges. Here, we use the same sample of simulated MBH merger host galaxies to investigate their stellar kinematics, as probed by optical integral field unit (IFU) spectroscopy. We perform stellar population synthesis and dust radiative transfer to generate synthetic 3D optical spectral datacubes of each simulated galaxy, and produce mock stellar kinematic maps. Based on a linear discriminant analysis of a combination of kinematic parameters derived from these maps, we show that this approach can identify MBH binary and merger host galaxies with accuracies that increase with chirp mass and mass ratio. For mergers with high chirp masses ($\gtrsim$10$^{8.2}$ $M_\odot$) and high mass ratios ($\gtrsim$0.5), the accuracies reach $\gtrsim$85\%, and their host galaxies are uniquely characterized by slower rotation and stronger stellar kinematic misalignments. These kinematic properties are commonly associated with massive early-type galaxies that have experienced major mergers, and naturally act as signposts for MBH binaries and mergers with high chirp masses and mass ratios. These results suggest that IFU spectroscopy should also play a role in telescope follow-up of future MBH binaries and mergers detected in gravitational waves. 

\end{abstract}


%%%%%%%%%%%%%%%% MAIN CONTENT BEGINS %%%%%%%%%%%%%%%% 

\section{Introduction} \label{sec:intro}

Mergers of massive black holes (MBHs, with masses of $M_\mathrm{BH} \gtrsim 10^5 M_\odot$) at the centers of galaxies are inevitable in our standard picture of hierarchical galaxy formation \citep{Begelman_1980, Volonteri_2003, Volonteri_2009}. Following a merger of two galaxies, their two central MBHs will sink to the gravitational potential minimum due to dynamical friction, and form a gravitationally-bound MBH binary \citep{Colpi_1999, Yu_2002, Pfister_2017}. This binary will then harden in the nuclear region of the galaxy, due to some combination of three-body interactions with stars \citep[e.g.,][]{Quinlan_1996, Merritt_2004, Berczik_2006, Vasiliev_2015, Sesana_2015} or torques from gas \citep[e.g.,][]{Armitage_2002, Escala_2004, Dotti_2007, Mayer_2007, Cuadra_2009, Chapon_2013, Fontecilla_2019}. Eventually, angular momentum loss becomes dominated by gravitational wave emission, which drives the MBH binary to coalescence \citep{Wyithe_2003, Sesana_2004}. 

MBH binaries and mergers are already beginning to be detected in low-frequency gravitational waves. Current pulsar timing array \citep[PTA;][]{Jenet_2004} experiments such as the North American Nanohertz Observatory for Gravitational Waves (NANOGrav), the European Pulsar Timing Array (ETPA), the Parkes Pulsar Timing Array (PPTA), and the Chinese Pulsar Timing Array (CPTA) have recently detected the nHz stochastic gravitational wave background \citep{nanograv_background, EPTA_background, ppta_background, cpta_background}, consistent with being produced by the ensemble population of MBH \emph{binaries} in nearby galaxies \citep{nanograv_smbh_population, EPTA_smbh_population, ppta_mbh_population}. Although these experiments have not yet detected an individual MBH binary system \citep{nanograv_individ_limit, nanograv_anisotropy, EPTA_cont_source}, forecasts predict that they will soon achieve sufficient sensitivity for a first detection \citep{Sesana2009, Ravi_2015, Mingarelli_2017, Kelley_2018}. Future mHz gravitational wave experiments such as the \emph{Laser Interferometer Space Antenna} \citep[\emph{LISA};][]{Amaro-Seoane_2017} will detect \emph{mergers} of MBH with mass $ 10^4 M_\odot \lesssim M_\mathrm{BH} \lesssim 10^8 M_\odot$, to high redshifts of $\gtrsim$10 \citep{Sesana_2005, Banks_2022}. These discoveries will represent a breakthrough in our ability to probe the formation and growth of MBHs over cosmic time.

Once an individual MBH binary or merger is detected in gravitational waves, identifying its host galaxy will be challenging. Identifying the exact host galaxy of each MBH binary or merger will enable a variety of key science goals, such as unveiling the structure of accretion flows around binary MBH \citep[e.g.,][]{Armitage_2002, Cuadra_2009, Noble_2012, Duffell_2020}, constraining cosmological parameters \citep{Schutz_1986, Holz_2005}, and probing the co-evolution of galaxies and their central MBHs \citep{Magorrian_1998, Ferrarese_2000, Gebhardt_2000}, which will constitute multi-messenger source detections. However, gravitational wave detections will have relatively poor sky localizations that contain an overwhelmingly large number of possible host galaxies in follow-up telescope observations. \citet{Goldstein_2019} and \citet{Petrov_2024} estimate that the error volumes of individual MBH binaries detected by PTAs will contain of order $\sim$10$^2$ candidate host galaxies, after both redshift and mass cuts (assuming empirical galaxy scaling relations) based on the gravitational wave information. Similarly, \citet{Lops_2023} show that the error volumes of MBH mergers detected by \emph{LISA} will contain of order $\sim$10$^{2-3}$ candidate host galaxies, for mergers of relatively-massive MBHs ($\gtrsim$10$^{7}$~$M_\odot$) at relatively-low redshifts ($z \lesssim 2$), which are most amenable to telescope follow-up. Clearly, additional approaches to selecting the best sources for follow-up will be needed to identify the exact host.

Many strategies identifying the host galaxies of MBH binaries and mergers have been suggested, although secure methods are still lacking. Previous studies have often focused on transient or time-variable electromagnetic signatures from gas surrounding MBH binaries \citep[e.g.,][]{Dorazio_2015, Graham_2015, Kelley_2019, Charisi_2022, Milosavljevic_2005, Megevand_2009, ONeill_2009, Corrales_2010, Rossi_2010} that can be observed in telescope follow-up \citep[see recent reviews by][]{Bogdanovic_2022, Dorazio_2023}. However, these time-domain signatures are highly uncertain, and may be applicable to only a small fraction of MBH binaries and mergers if the majority of their surrounding environments are too gas-poor to support accretion \citep{Izquierdo-Villalba_2023a, Dong-Paez_2023}, or are heavily enshrouded in dust that obscures the electromagnetic emission from the accretion flow \citep{Koss_2018}. 

Recently in \citet[][hereafter Paper I]{Bardati_2024}, we developed an alternative approach to identifying the host galaxies of MBH binaries and mergers based on galaxy morphology in telescope imaging, thus demonstrating that the host galaxy stellar emission contain unique signatures. Specifically, we used simulated MBH merger host galaxies from a cosmological simulation, and performed dust radiative transfer to produce synthetic broadband ultraviolet, optical, and infrared images. Using a variety of morphological parameters extracted from these synthetic images, we trained a Linear Discriminant Analysis (LDA) predictor to distinguish between MBH merger host galaxies and a control sample. We showed that the accuracy of this method increases with both chirp mass and mass ratio of the MBH binary, reaching $\gtrsim$80\% for mergers with high chirp masses ($\gtrsim$10$^{8.2}$ $M_\odot$) and high mass ratios ($\gtrsim$0.5). We showed that this is because the primary signatures of these host galaxies are prominent classical bulges, which are built from major mergers of massive galaxies; these galaxy mergers naturally lead to the formation of MBH binaries and mergers with high mass ratios and high chirp masses. These results are broadly consistent with other studies that use cosmological simulations to show that the host galaxies of MBH binaries detectable by PTA experiments (which preferentially have high mass ratios and high chirp masses) are massive early-type galaxies with low star formation rates and high metallicities \citep[e.g.,][]{Saeedzadeh_2024, Cella_2024}, while the host galaxies of MBH mergers detectable by LISA (which have low chirp masses) are not morphologically distinct \citep{Izquierdo-Villalba_2023b}. Although our results from Paper I imply that there are indeed observable signatures of MBH binaries and mergers with high mass ratios and high chirp masses in their host galaxy starlight, it is unclear whether unique host galaxy morphologies in imaging is actually the most robust observational signature.

Here, we investigate the unique signatures of MBH binaries and mergers in their host galaxy \emph{stellar kinematics} (in contrast to morphology in Paper I), as probed by optical integral field unit (IFU) spectroscopy with medium spectral resolution. Although IFU spectroscopy is significantly more resource-intensive than imaging, a stellar kinematics approach may have unique advantages over morphology. For example, if the primary distinguishing characteristic of MBH merger host galaxies is indeed the presence of a prominent bulge, then this feature will likely be more distinctive in stellar kinematic maps than morphology in imaging. Furthermore, \citet{Nevin_2021} use simulations to show that stellar kinematic signatures of galaxy mergers persist for several Gyrs, after morphological disturbances have disappeared and when the MBH binary or merger would be detected in gravitational waves. Thus, it is possible that the stellar kinematics of MBH binary and merger host galaxies can contain unique signatures from \emph{both} permanent features such as bulges, as well as kinematic disturbances from their preceding galaxy merger. %However, we emphasize that our approach is \emph{not} predicated on using kinematic signatures of galaxy mergers as signposts of MBH mergers. Similar to \citet{Bardati_2024}, we investigate signatures of MBH mergers from their host galaxy stellar kinematics in an agnostic way, and are thus sensitive to both permanent and transient kinematic signatures. 

To assess whether stellar kinematics can identify MBH binary and merger host galaxies, we use a sample of 201 simulated MBH merger host galaxies from a cosmological simulation, as well as a mass- and redshift-matched control sample for comparison. The samples we use here are the exact same samples we previously used in Paper I. However, the post-processing and analysis of these simulated galaxies here are distinct from Paper I, since we are producing and analyzing synthetic 3D spectral datacubes instead of synthetic images. We perform stellar population synthesis and dust radiative transfer to generate synthetic spectral datacubes of each simulated galaxy, similar to observations from optical integral field unit (IFU) spectrographs. We then perform full-spectrum fitting of these spectral datacubes to produce 2D stellar kinematic maps, and extract a variety of kinematic parameters for each simulated galaxy. By training a LDA predictor on these kinematic parameters, we find a linear combination of kinematic parameters that can optimally distinguish between the MBH merger host galaxies and the control sample with high accuracies. Our results thus motivate follow-up IFU observations (or analysis of archival IFU data) of galaxies lying in the localization regions of MBH binaries and mergers detected in gravitational waves.

The outline of this paper is as follows. In Section~\ref{sec:sims}, we briefly describe the cosmological simulation we use, and our selection of samples of simulated galaxies. In Section~\ref{sec:radtrans}, we describe the radiative transfer simulations we perform to produce 3D spectral datacubes, as well our fitting of these datacubes to produce 2D stellar kinematic maps. In Section~\ref{sec:analysis}, we describe the various kinematic parameters we extract, and our linear discriminant analysis. In Section~\ref{sec:discussion}, we present our main results, and use them to build a consistent picture of the unique signatures of MBH binary and merger host galaxies. We briefly summarize and conclude in Section~\ref{sec:conclusions}.

%%%%%%%%%%%%%%%%%%%%%%%%%%%%%%%%%%%%%%%%%%%%%%%%%%%%%
\section{Cosmological Simulation \ and Galaxy Samples Selection} \label{sec:sims}
In this section, we describe the cosmological simulation, and our selection of samples of simulated galaxies. We note that the simulation and galaxy samples we use are identical to those used in \citet[][Paper I]{Bardati_2024}. This enables a direct comparison between previous results from the galaxy morphology approach in Paper I to our stellar kinematics approach here. We thus only briefly describe the most salient points below, and refer to Paper I for more details.

\subsection{Romulus25}\label{subsec:romulus}

For our analysis, we use the Romulus25 cosmological simulation of galaxy formation \citep{Tremmel_2017}. The simulation is run to $z=0$ using the N-Body + Smooth Particle Hydrodynamics code \texttt{ChaNGa} \citep{Menon_2015}, with a 25 cMpc per side volume and a standard $\Lambda \mathrm{CDM}$ cosmology consistent with \citet{Planck_Collaboration_2016}.

Romulus25 introduces new implementations of MBH formation, growth, and dynamics. MBHs in Romulus25 are seeded from direct collapse with masses of $M_{BH} = 10^6 M_\odot$. MBH formation is directly tied to the properties of the surrounding gas particles, such as metallicity, gas density, and temperature, which better reproduces MBH seeding at high redshift than methods tied to halo mass. After formation, the MBHs grow both via accretion of surrounding gas at a rate determined by a modified Bondi-Hoyle prescription, and via numerical mergers with other MBHs when their mutual distance is less than two softening lengths ($\lessapprox$700~pc). These MBH binaries are tracked to sub-kpc scales using a sub-grid model of dynamical friction \citep{Tremmel_2015}, in contrast to the more common approach of artificially pinning the MBHs to their host galaxy potential minimum at each timestep. This allows MBHs to decouple from their host galaxy dynamics, thus enabling MBH dynamics to be tracked down to $\lessapprox$700~pc before merging numerically below the resolution limit of the simulation. This feature of Romulus25 leads to shorter time-delays between the numerical merger and the physical merger, when gravitational waves would be emitted. We refer to \citet{Tremmel_2017} for more details on Romulus25. 

\citet{Jung_2022} investigated the properties of Brightest Group Galaxies (BGGs) in Romulus25, and found that their stellar kinematics are an excellent match to observations, especially for BGG in lower-mass groups with halo masses of $<$10$^{13}$~$M_\odot$. These lower-mass BGGs have halo masses similar to our subsample of MBH merger host galaxies (and the corresponding control galaxy sample) with chirp masses of $\gtrsim$10$^{8.2}$ $M_\odot$ that we highlight below in Sections~\ref{subsec:LDA} and \ref{sec:discussion}. Thus, the results of \citet{Jung_2022} suggests that the stellar kinematics of simulated galaxies from Romulus25 are realistic and robust.

\subsection{MBH Merger Host Galaxies Sample}\label{subsec:mergersampleselection}

We select a sample of 201 MBH numerical mergers in Romulus25, consisting of all MBH merger events that occur at a redshift of $z \leq 2$, with a mass ratio of $\nicefrac{M_2}{M_1} > 0.1$ (where $M_1 > M_2$) and with MBH masses of $M_{\text{BH}} \gtrsim 10^7 M_\odot$. These cuts select relatively massive and low-$z$ galaxies, suitable for follow-up telescope observations.

We use the \texttt{Amiga Halo Finder} \citep{Knollmann_2009} to identify the host galaxy of each MBH merger at the nearest available simulation snapshot in time, subsequent to the numerical merger. Since the time-delay between numerical and physical merger is poorly-understood, we also track these galaxies for $\sim$1 Gyr after their MBH merger to investigate the time evolution of our LDA results in Section~\ref{subsec:timescales} below. In Paper I, we showed that this sample of MBH merger host galaxies follows the typical scaling relation between central MBH mass ($M_{\text{BH}}$) and stellar mass ($M_{*}$), consistent with other studies which investigate MBH mergers using cosmological simulations \citep[e.g.,][]{DeGraf_2021}.

\subsection{Control Galaxy Sample}\label{subsec:controlsampleselection}

We select a mass- and redshift-matched control sample of galaxies in Romulus25 that do not host MBH mergers, for comparison to our MBH merger host sample. To produce this control sample, we first bin the simulation snapshot timesteps such that each time bin contains a minimum of 30 MBH mergers. Within each time bin, we then bin the MBH host galaxies by mass. In each of these mass bins, we select galaxies which are not in the MBH merger host galaxies sample, weighted by the ratio of the number of MBH merger galaxies to the number of galaxies that are not MBH merger galaxies. This produces a sample of non-MBH merger sample galaxies that have a mass distribution similar to the MBH merger sample in each time bin. The collection of time-binned samples is the resulting mass- and redshift-matched control sample, thereby avoiding biases in our results stemming from galaxy mass and redshift. We emphasize that our use of a cosmological simulation for this investigation naturally takes into account the diverse merger histories of galaxies in both the MBH merger host galaxies and control galaxies sample, since these simulated galaxies are grown in a realistic cosmological context.

%%%%%%%%%%%%%%%%%%%%%%%%%%%%%%%%%%%%%%%%%%%%%%%%%%%%%
\begin{figure*}[t!]
\centering
\includegraphics[width=0.9\textwidth]{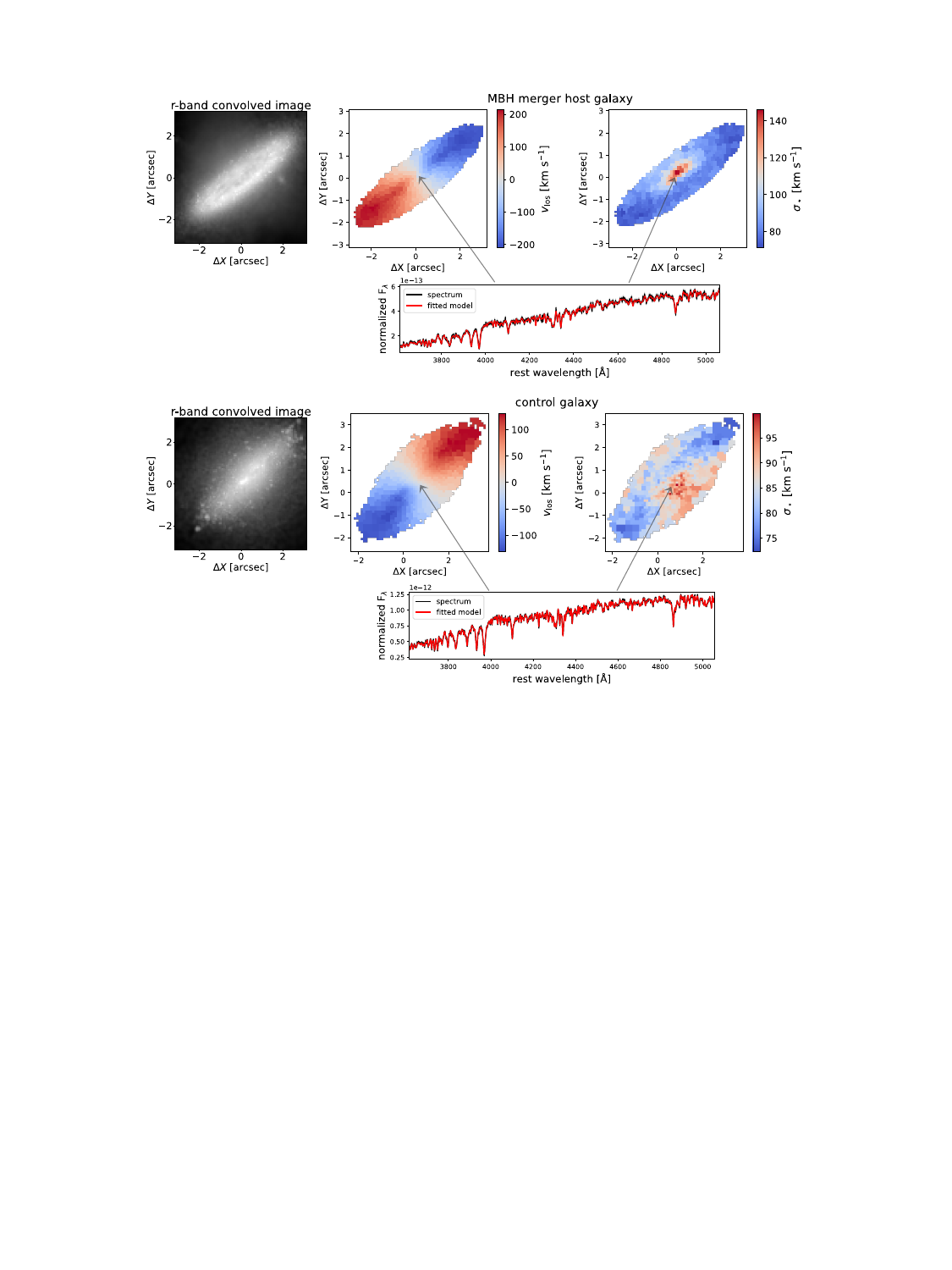}
\figcaption{Stellar kinematic maps of a representative simulated galaxy from our MBH merger hosts sample (top panels), as well as from our control sample (bottom). The top left panel for each of the two galaxies displays the stellar line of sight velocity ($v_\mathrm{los}$) and the top right panel shows the stellar velocity dispersion ($\sigma_\star$). The bottom panels show an example synthetic spectrum from the indicated spaxel in the datacube (black line), as well as the fitted model from \texttt{ppXF} (red line). It is difficult to visually classify these galaxies from these kinematic maps, and thus we extract a variety of kinematic parameters in Section \ref{subsec:mergerpredictors} to input into a linear discriminant analysis predictor.}
%% Future reference: halos are 6400_39_2 and 6350_61_1
\label{fig:kinematic_maps}
\end{figure*}
%%%%%%%%%%%%%%%%%%%%%%%%%%%%%%%%%%%%%%%%%%%%%%%%%%%%%

%%%%%%%%%%%%%%%%%%%%%%%%%%%%%%%%%%%%%%%%%%%%%%%%%%%%%
\section{Generating 3D Spectral Datacubes \ and Stellar Kinematic Maps}\label{sec:radtrans}

In this section, we describe the dust radiative transfer simulation we perform to generate synthetic 3D spectral datacubes of the simulated galaxies from Sections~\ref{subsec:mergersampleselection} and~\ref{subsec:controlsampleselection} above, and our fitting of these datacubes to produce mock stellar kinematic maps. We emphasize that in both these steps, we do not attempt to carefully match the specifications of any particular telescope or instrument. Instead, our work here is a first step to gauge the efficacy of IFU spectroscopy for identifying MBH merger host galaxies, and we aim to make our results broadly applicable to a wide range of instruments. Our inherent assumptions in generating the spectral datacubes (e.g., medium-resolution spectroscopy and nearly diffraction-limited seeing with $0\farcs1$ pixel scales) are already viable with current IFU instruments such as the Multi Unit Spectroscopic Explorer \citep[MUSE;][]{Bacon_10} on the Very Large Telescope, and achievable with relatively short exposure times using future instruments such as  High Angular Resolution Monolithic Optical and Near-infrared Instrument \citep[HARMONI;][]{Thatte_20} on the Extremely Large Telescope. We discuss some practical considerations for applying our results to new and archival IFU observations in Section~\ref{subsec:observations}, and defer a more careful matching to instrument specification (e.g., using the \texttt{HSIM3} simulator for HARMONI observations; \citealt{Zieleniewski_15}) for future work.

\subsection{Dust Radiative Transfer with SKIRT}\label{subsec:SKIRT}
We use the \texttt{SKIRT} 3D Monte Carlo dust radiative transfer software \citep{Baes_2011, Camps_2015} to generate synthetic spectral datacubes for each simulated galaxy. 
\texttt{SKIRT} performs stellar population synthesis on each star particle of a simulated galaxy, and uses the resulting SEDs to propagate photon packets through the galaxy's interstellar medium to produce spatially-resolved galaxy SEDs. Critically, the current version 9 of \texttt{SKIRT} includes stellar and gas kinematics in the radiative transport by accounting for Doppler shifts due to particle motion in the photon packets \citep{Camps_20}. This enables us to accurately generate line-of-sight velocity dispersion maps in Section~\ref{subsec:ppxf}.

To first perform the stellar population synthesis, we
use the Flexible Stellar Population Synthesis \citep[\texttt{FSPS};][]{Conroy_Gunn_2010} package included in \texttt{SKIRT}, which models each star particle as a single-age simple stellar population. In \texttt{FSPS}, we assume a \citet{Kroupa_2001} initial mass function, and we use the MIST evolutionary tracks \citep{Jieun_2016}. To produce a spectrum from the single stellar population of each star particle, we use the MILES \citep{Sanchez_2006} spectral libraries \citep{Vazdekis_10}. Since we are interested primarily in the stellar kinematics, we neglect stellar and nebular emission lines, and we do not include emission from an active galactic nucleus (AGN). Luminous AGN would appear in only a small fraction of our simulated galaxies, and in practice their emission would be removed in the full-spectrum fitting to produce kinematic maps, by including template AGN spectra in the spectral fitting.

For \texttt{SKIRT} to perform dust radiative transport, the properties of the dust must be modeled. We thus assume that the spatial distribution of the dust follows that of the gas particles in each galaxy. We use a THEMIS dust mix \citep{Jones_2017} to represent our dust grain size and composition, and assume a proportionality factor of 0.2 on the grain size distribution function. In our tests, we find that our results are not strongly dependent on our choice of these dust parameters.

We use \texttt{SKIRT} to produce output spectra over the rest-frame wavelength range of 3400 -- 7000~\AA, with a spectral resolution of $R = 4000$. This wavelength range includes prominent stellar absorption lines (such as the H Balmer series and the Ca II H and K lines) that are commonly used to produce stellar light-of-sight velocity and velocity dispersion maps in IFU observations. These spectra also enable us to produce synthetic $u$-, $g$-, and $r$-band images, which are used in our calculation of some of the kinematic parameters described in Section~ \ref{subsec:mergerpredictors}. For our synthetic spectral datacubes, we assume a pixel scale of $0\farcs1$ pixel$^{-1}$ to match the specifications used in Paper I for direct comparison. At the median redshift ($z \sim 1$) of the simulated galaxies in our sample, each pixel corresponds to a physical scale of approximately 0.8~kpc, larger than the gravitational softening scale of Romulus25.

We compute uncertainties on the flux density values in each synthetic spectrum using reliability statistics output from \texttt{SKIRT}, following prescriptions from \citet{Camps_2018} and \citet{Camps_20}. In addition to measuring the accumulated weights $w_i$ of the $N$ individual photon packets launched by the Monte Carlo radiative transfer simulation in each pixel $\sum_i^N w_i$, we also track the higher-order moments $W_k \equiv \sum_i^N w_i^k$ for $k\in\{2, 3, 4\}$. With these moments, we compute the relative error
\begin{equation*}
R = \sqrt{\frac{W_2}{W_1^2} - \frac{1}{N}},
\end{equation*}
as well as the variance of the variance 
\begin{equation*}
VOV = \frac{1}{R^4}\left(\frac{W_4}{W_1^4} - \frac{4W_3}{W_1^3 N} + \frac{6W_2}{W_1^2N^2} - \frac{3}{N^3}\right) - \frac{1}{N},
%VOV = \frac{W_4 - 4W_1W_3/N + 8W_2W_1^2/N^2 - 4W_1^4/N^3 - W_2^2/N}{(W_2 - W_1^2/N)^2},
\end{equation*}
which is a measure of the uncertainty in $R$. We use these quantities as a measure of the uncertainty on the flux in each pixel, and to mask bad pixels (see Section \ref{subsec:ppxf}). 

To minimize the effects of galaxy orientation on our results, we generate spectral datacubes along four different isotropically-oriented viewing angles for each galaxy. Specifically, for each simulated galaxy, we generate spectral datacubes from the perspective of each of the four vertices of a regular tetrahedron. Since each galaxy is randomly oriented, the tetrahedron is also effectively randomly oriented, and thus the resultant datacubes sample a variety of different viewing angles. 

\subsection{Producing Stellar Kinematic Maps with ppXF}\label{subsec:ppxf}

To produce a stellar kinematics map of each galaxy, we fit its spectral datacube using the penalized-pixel fitting software \texttt{ppXF} \citep{Cappellari_23}. \texttt{ppXF} uses single stellar population templates to perform full-spectrum fitting of galaxy spectra. The stellar templates are convolved with a line-of-sight velocity distribution to fit kinematic properties from the observed absorption lines. To include the effects of a Point Spread Function (PSF), we convolve the monochromatic image at each wavelength slice of the spectral datacubes with a $0\farcs1$ full-width at half-maximum (FWHM) Gaussian. Since we are not directly comparing our synthetic spectra datacubes to observed data, we do not include the effects of a line-spread function. Before fitting, we mask out all spectra in each datacube that have $>$10\% of pixels with relative error $R >0.1$ and variance of variance $VOV > 0.1$; SKIRT spectra satisfying these conditions are known to be of insufficient quality and unreliable for fitting \citep{Camps_2015}.

For each datacube, we fit the unmasked spectra in Voronoi bins to obtain a line-of-sight velocity $v_\mathrm{los}$ and velocity dispersion $\sigma_\star$ for the stellar component. Following observational approaches, we first Voronoi bin each spectral datacube using \texttt{VorBin} \citep{Cappellari_03} to achieve a minimum signal-to-noise ratio of 50 in each bin at a fiducial wavelength of 5000~\AA. We then use \texttt{ppxf} to fit the combined spectra in each Voronoi bin, while masking out the pixels in each combined spectrum that have relative error $R >0.1$ and variance of variance $VOV > 0.1$. Since our \texttt{SKIRT} spectra do not include nebular emission, we do not include any emission line components in our spectral fitting. We fit the spectra over the 3450 -- 6000~\AA~wavelength range, using the empirical MILES \citep{Sanchez_2006} library of single stellar population template spectra \citep{Vazdekis_10}, and include the two-parameter dust model from \citet{Cappellari_23}. In our fits, we include the 1$\sigma$ uncertainties on the flux densities in the spectra from photon noise in the \texttt{SKIRT} radiative transfer. We correct the final $\sigma_\star$ values for differences in spectral resolution between the templates and simulated spectra, following \citet{Cappellari_17}.

Figure~\ref{fig:kinematic_maps} shows the resulting $v_\mathrm{los}$ and $\sigma_\star$ stellar kinematic maps of an example galaxy from our MBH merger sample, as well as a similar galaxy from the control sample. Figure~\ref{fig:kinematic_maps} also displays an example spectrum from a central spaxel in both datacubes, along with the \texttt{ppXF} fit. In all of the Voronoi bins, the full-spectrum fits are robust, with $\chi^2/dof \approx 1$. We will use these these stellar kinematic maps to extract a variety of kinematic parameters in Section~\ref{sec:analysis}, and use a combination of these parameters to identify MBH merger host galaxies.

%%%%%%%%%%%%%%%%%%%%%%%%%%%%%%%%%%%%%%%%%%%%%%%%%%%%%
\begin{figure*}[ht]
\centering
\includegraphics[width=0.75\textwidth]{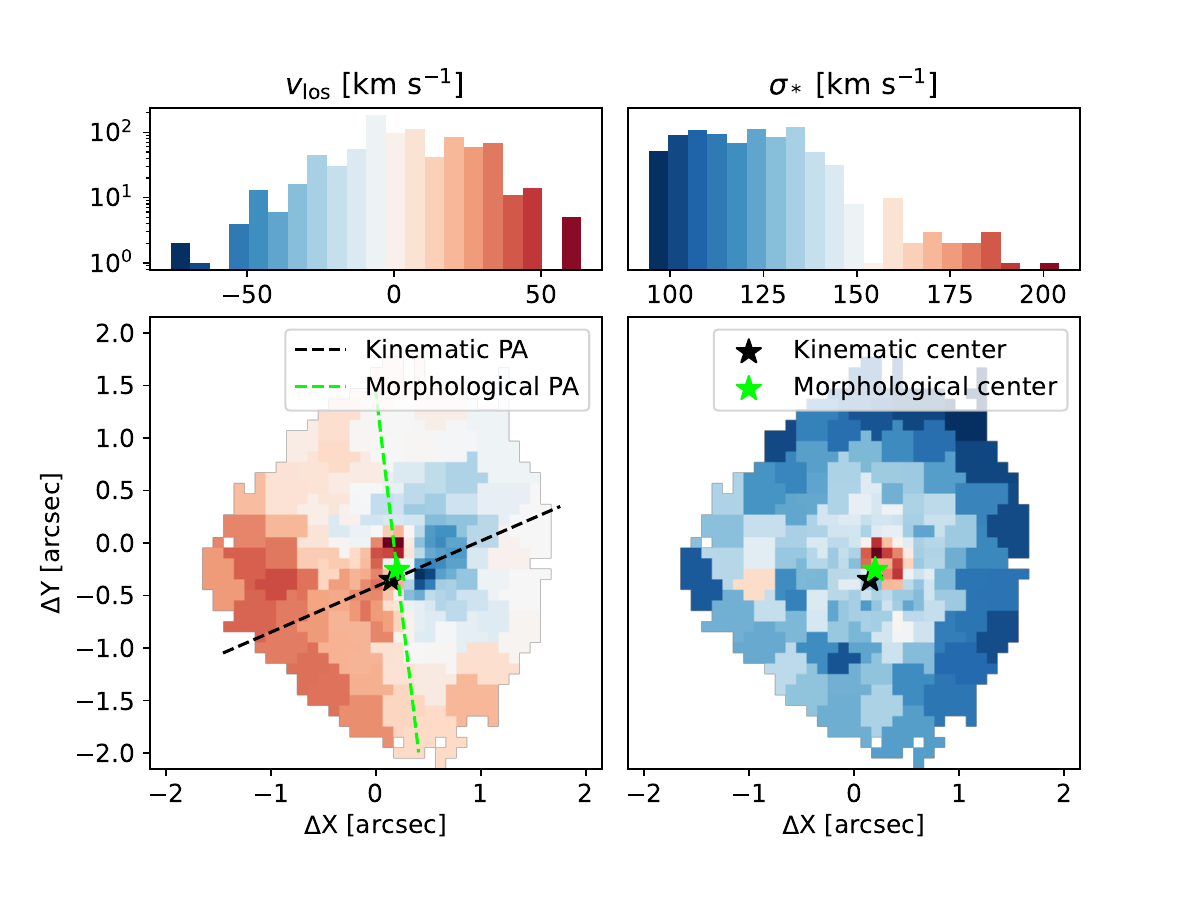}
\figcaption{Our fitting of the line of sight velocity $v_\mathrm{los}$ map (left) and stellar velocity dispersion $\sigma_\star$ map (right) to extract stellar kinematic properties for the example MBH merger host galaxy from Figure~\ref{fig:kinematic_maps}. The bottom left panel shows the $v_\mathrm{los}$ map produced by \texttt{ppXF}, along with the kinematic position angle fitted using the $v_\mathrm{los}$ map (blue dashed line), as well as the photometric position angle fitted using the synthetic $r$-band image (red dashed line). The difference between these two position angles is used to calculate the $\Delta$PA parameter. The bottom right panel shows the $\sigma_\star$ map produced by \texttt{ppXF}, along with the kinematic center fitted using the $\sigma_\star$ map (blue cross) and morphological center fitted using the synthetic $r$-band image (red cross). The top left and top right panels show histograms of the distributions of $v_\mathrm{los}$ and $\sigma_\star$ values in the kinematic maps, respectively, which are used to compute the statistical moments $\mu_v$ and $\mu_\sigma$ described in Section~\ref{subsec:mergerpredictors}.
}
\label{fig:kinematic_parameter_extraction}
\end{figure*}

% things to potentially also include in plot:
%   - morphology calculations on sigma map?
%   - put spin param somewhere?
% 
%%%%%%%%%%%%%%%%%%%%%%%%%%%%%%%%%%%%%%%%%%%%%%%%%%%%%

%%%%%%%%%%%%%%%%%%%%%%%%%%%%%%%%%%%%%%%%%%%%%%%%%%%%%

\section{Analysis of the Stellar Kinematic Maps}\label{sec:analysis}

\subsection{Extracting Kinematic Parameters}\label{subsec:mergerpredictors}

From the kinematic maps of $v_\mathrm{los}$ and $\sigma_\star$ produced by \texttt{ppxf}, we extract 12 stellar kinematic measurements, described in detail below. To calculate these kinematic parameters, we follow the definitions from \citet{Nevin_2021}, and use the Radon Transform from \citet{Stark_2018} and the \texttt{kinemetry} package \citep{Krajnovic_2006}. Some of these kinematic parameters also use parameters from broadband images, such as the $r$-band effective radius and ellipticity $\varepsilon$. We thus also use our spectral datacubes to produce synthetic $r$-band images, and use the \texttt{StatMorph} package \citep{Rodriguez-Gomez_2019} to compute these parameters. 

We first calculate the statistical moments of the line-of-sight velocity ($\mu_{2, v}$, $\mu_{3, v}$, $\mu_{4, v}$) and velocity dispersion maps ($\mu_{1, \sigma}$, $\mu_{2, \sigma}$, $\mu_{3, \sigma}$, $\mu_{4, \sigma}$). The subscripts 1 through 4 denote the mean, dispersion, skew, and kurtosis of the 1-dimensional distributions. We use the mean line-of-sight velocity ($\mu_{1, v}$) to normalize the line-of-sight velocity maps, and thus it is not included as a parameter in the LDA. Since we found in Paper I that MBH merger host galaxies have more prominent bulges in imaging, we expect their mean velocity dispersion ($\mu_{1, \sigma}$) to be higher.

To compute the kinematic asymmetry parameters $A$ and $A_2$, as well as the kinematic center (used in our other kinematic measurements), we follow the technique described in \citet{Nevin_2021}. Specifically, we use the bounded Absolute Radon Transform from \citet{Stark_2018}, defined as 
\begin{equation*}
R_\mathrm{AB}(r, \theta) = \int_0^{R_e} |v(x, y) - \langle v(x, y) \rangle|dl,
\end{equation*}
where for a given polar coordinate $(r, \theta)$, the line-of-sight velocity map $v(x, y)$ is integrated over the line segment passing through $(r, \theta)$ and perpendicular to the vector from the origin to $(r, \theta)$, subtracted by the mean line-of-sight velocity along that line segment $\langle v(x, y)\rangle$. The line segment is bounded to a width that we set to the $r$-band effective (i.e., half-light) radius $R_e$, as calculated by \texttt{StatMorph}. Since the Radon Transform is sensitive to the location of the origin used, we follow the method described by \citet{Stark_2018} to find the kinematic center. We use this kinematic center in all further kinematic calculations that require a center point. We also use the Radon Transform calculation to find the asymmetry parameters $A$ and $A_2$, as defined by \citet{Nevin_2021}, but these asymmetry parameters are excluded from the LDA because we find that they are not significantly different between the MBH merger and control galaxy samples.

We use \texttt{kinemetry} \citep{Krajnovic_2006} to calculate the difference between photometric and kinematic position angles $\Delta$PA from the line-of-sight velocity map, centered around the kinematic center calculated earlier. The kinematic position angle $\mathrm{PA}_\mathrm{kin}$ is calculated by \texttt{kinemetry} by modelling the line-of-sight velocity map $v(x, y)$ as a series of elliptical rings, whose finite harmonic expansion is 
\begin{equation*}
v(r, \theta) \approx A_0(r) + \sum_{n=1}^N \left[A_n(r) \sin(n\theta) + B_n(r) \cos(n\theta)\right]. 
\end{equation*}
The best fit model is then found by minimizing $\chi^2$ with respect to the parameters $A_n$ and $B_n$. We then compare the resulting kinematic position angle $\mathrm{PA}_\mathrm{kin}$ from the best-fit model to the position angle $\mathrm{PA}_\mathrm{img}$ calculated from the $r$-band image (using \texttt{StatMorph}) to obtain $\Delta \mathrm{PA} \equiv |\mathrm{PA}_\mathrm{kin} - \mathrm{PA}_\mathrm{img}|$. Galaxies with larger position angle differences $\Delta$PA display stronger stellar kinematic misalignments. We also calculate the residual between the model and actual galaxy velocity maps ($\mathrm{resid}$) as defined by \citet{Nevin_2021}, but ultimately exclude it since it does not significantly discriminate between the MBH merger and control galaxy samples.

We also calculate the spin parameter $\lambda_{R_e}$ \citep{Emsellem_2007}, which measures the specific angular momentum of the galaxy. We define it as 
\begin{equation*}
\lambda_{R_e} = \frac{\sum_{n=1}^N F_n r_n |v_n|}{\sum_{n=1}^N F_n r_n \sqrt{v_n^2+\sigma_n^2}},
\end{equation*}
where $F_n$ is the $r$-band flux, $r_n$ is the radius to the kinematic center, $v_n$ is the line-of-sight velocity map, $\sigma_n$ is the velocity dispersion map, and $N$ is the number of pixels. We also compute the morphological ellipticity $\varepsilon$ from the $r$-band image (using \texttt{StatMorph}) to compute the slow rotator discriminant, defined by \citet{Cappellari_2016} as 
\begin{equation*}
\lambda_{R_e} < 0.08 + \varepsilon/4 \;\;\; \mathrm{and} \;\;\; \varepsilon < 0.4.
\end{equation*}
Slow rotator galaxies (i.e., with small $\lambda_{R_e}$) are observed to be dispersion-dominated, and are more likely to have disturbed kinematics \citep{Emsellem_2011}. 

% We calculate the estimated proper distance between the kinematic center and imaging center in the $r$-band. ...(velocity with kinematic center above and sigma with low-pass filter, see Nevin) ...  We do not notice a significant difference between the MBH merger host and samples with these parameters and thus do not include them in the LDA. We note that this measurement is limited by spatial resolution, and thus may be more relevant for lower $z$ galaxies than probed here. ---> I think I will just exclude these parameters from the paper entirely.

Finally, we calculate morphological properties ($Gini_\sigma$, $M_{20, \sigma}$, $C_\sigma$, $A_\sigma$, $S_\sigma$, $A_{S, \sigma}$) directly on the $\sigma_\star$ maps. We use the same calculations as in Paper I, which are based on the definitions from \citet{Lotz_2004} and \citet{Pawlik_2016}.  

%Finally, we calculate morphological properties ($Gini$, $M_{20}$, $C$, $A$, $S$, $A_S$) on the $\sigma_\star$ and $v_\mathrm{los}$ maps (in addition to the synthetic $r$-band images). We use the same calculations as in \citet{Bardati_2024}, which are based on the definitions from \citet{Lotz_2004} and \citet{Pawlik_2016}. These calculations are easily directly applied to the $\sigma_\star$ map since it is positive-definite. For the $v_\mathrm{los}$ map, we take the absolute value to ensure the the $v_\mathrm{los}$ values are positive. These parameters do exhibit differences between the MBH merger host galaxies and the control galaxies, such as higher concentration in the $\sigma_\star$ of high mass and mass-ratio merger host galaxies when compared to control galaxies. However, the trained LDA in Section~\ref{subsec:LDA} does not select these parameters for inclusion due to their strong correlation to other parameters such as $\mu_{1, \sigma}$ or $\Delta$PA. 

\begin{figure*}[ht]
\centering
    \includegraphics[width=0.335\textwidth, keepaspectratio]{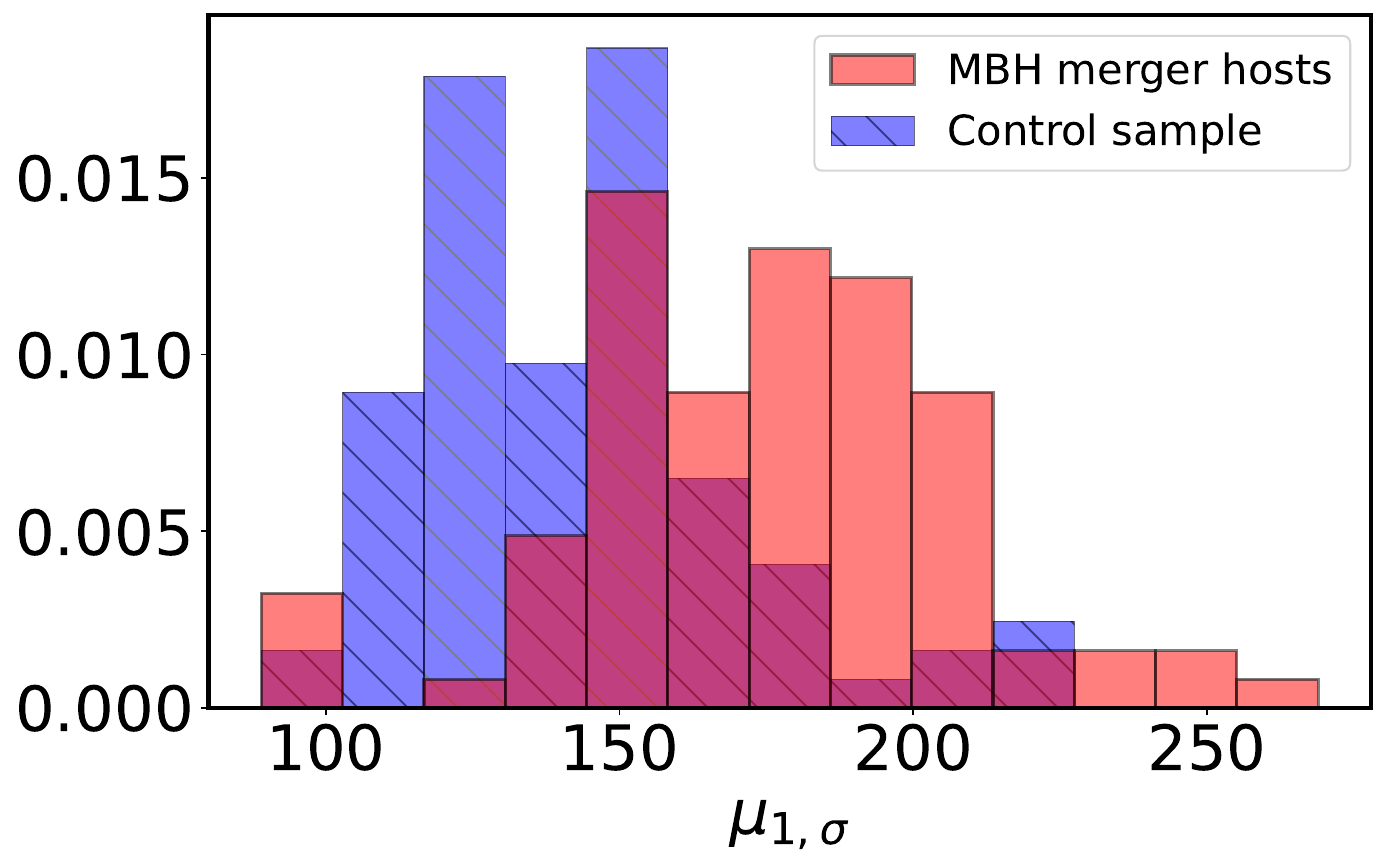} %width=0.32
\hspace{5pt}
    \includegraphics[width=0.31\textwidth, keepaspectratio]{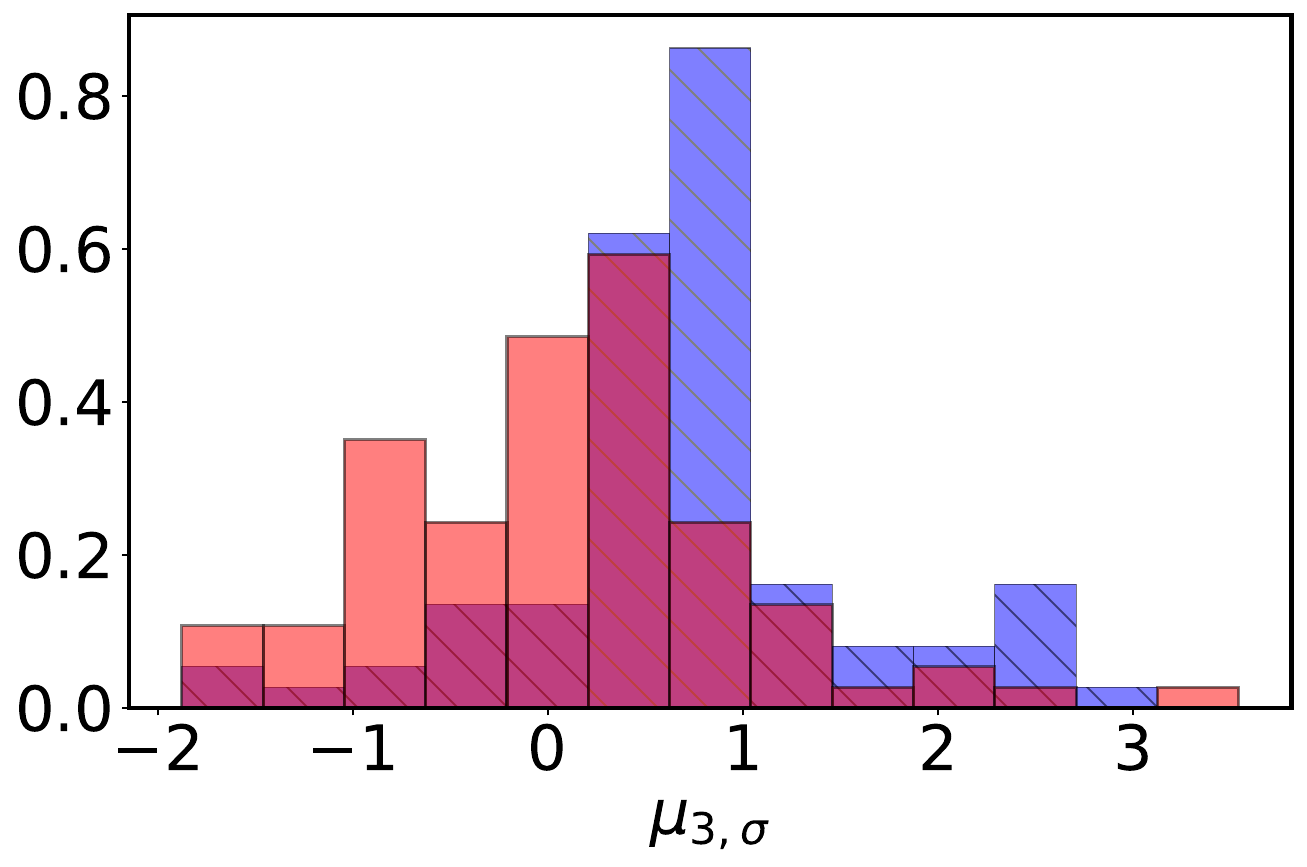}\\  %width=0.33\textheight
\vspace{10pt} 
    \includegraphics[width=0.305\textwidth, keepaspectratio]{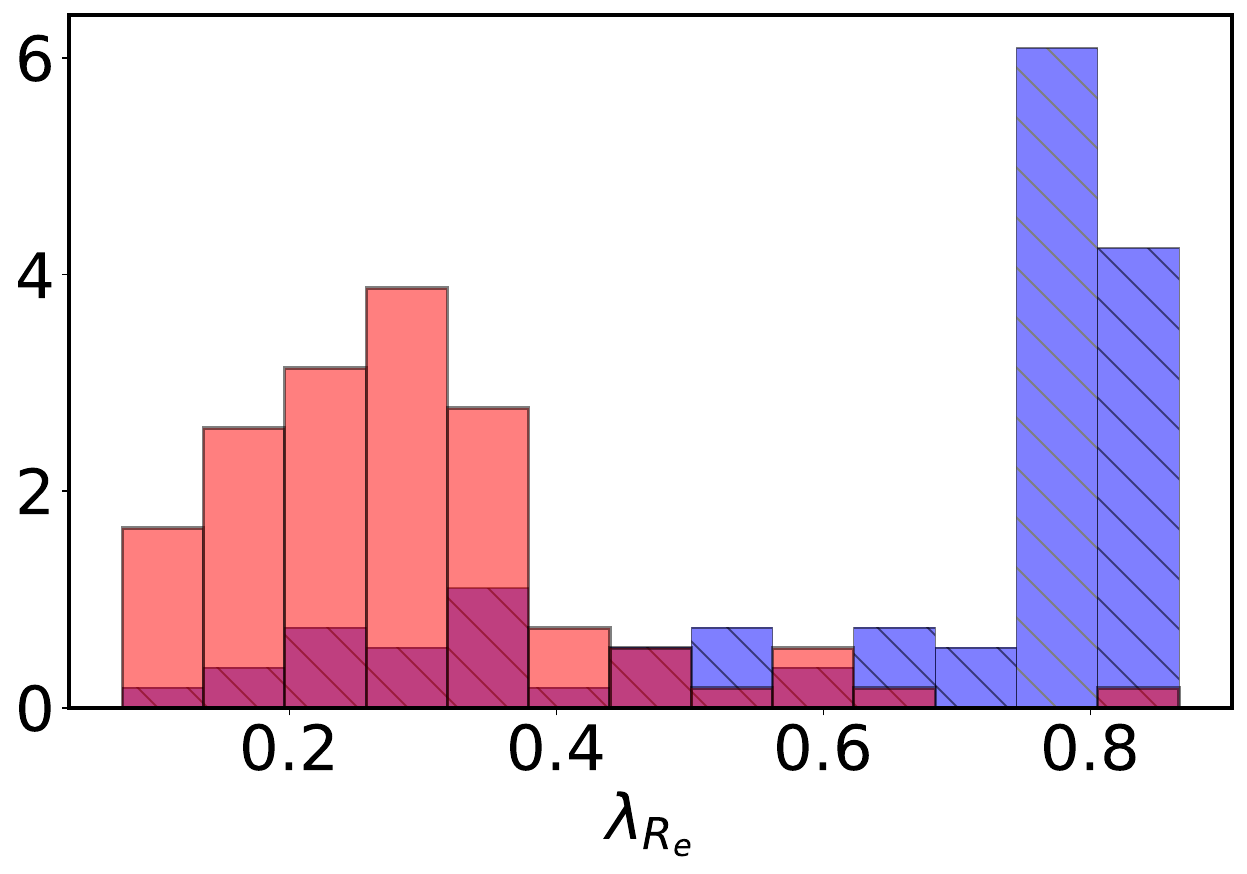} %width=0.33
\hspace{5pt}
    \includegraphics[width=0.32\textwidth, keepaspectratio]{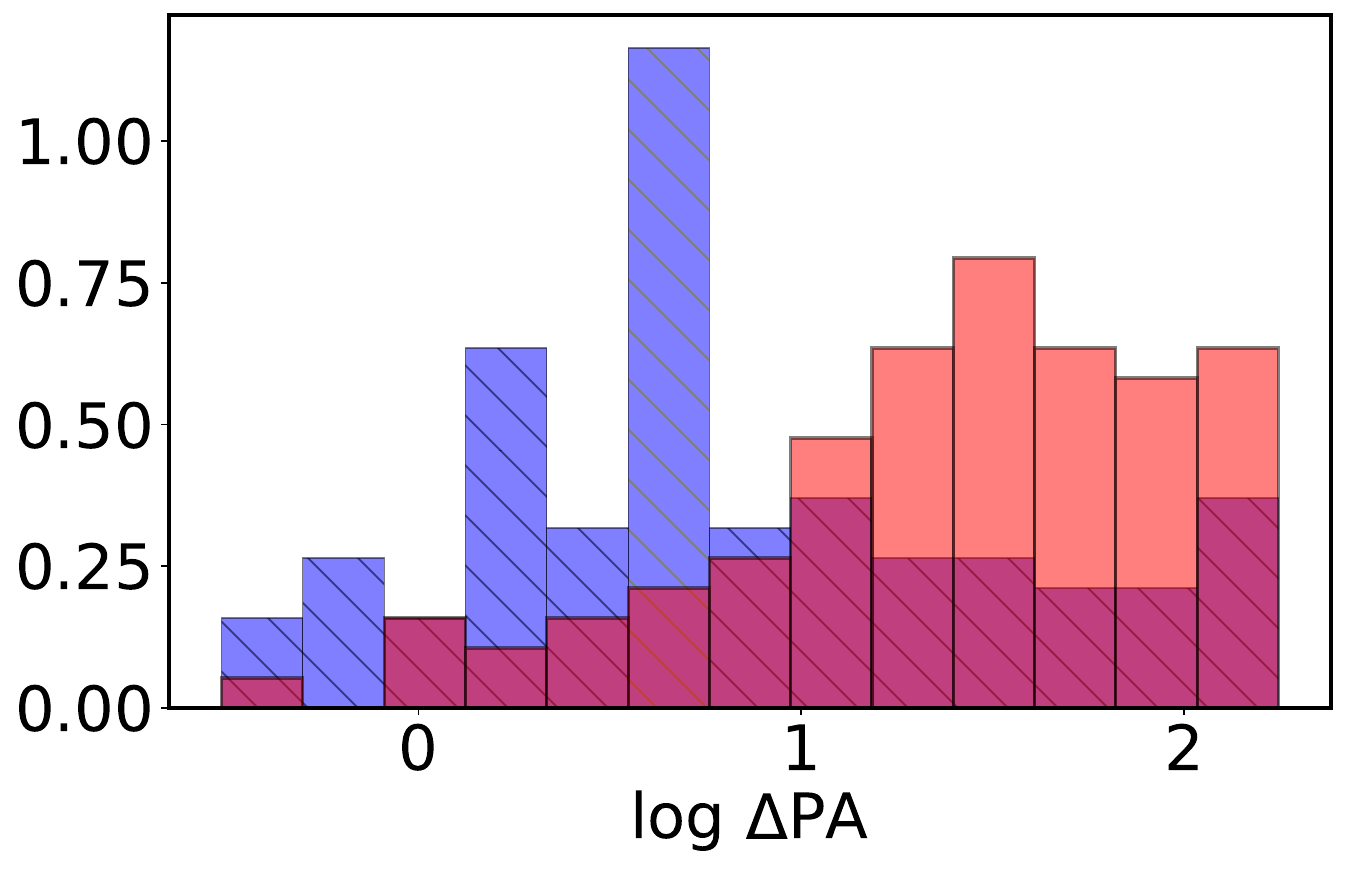} %width=0.32
\hspace{5pt}
    \includegraphics[width=0.3075\textwidth, keepaspectratio]{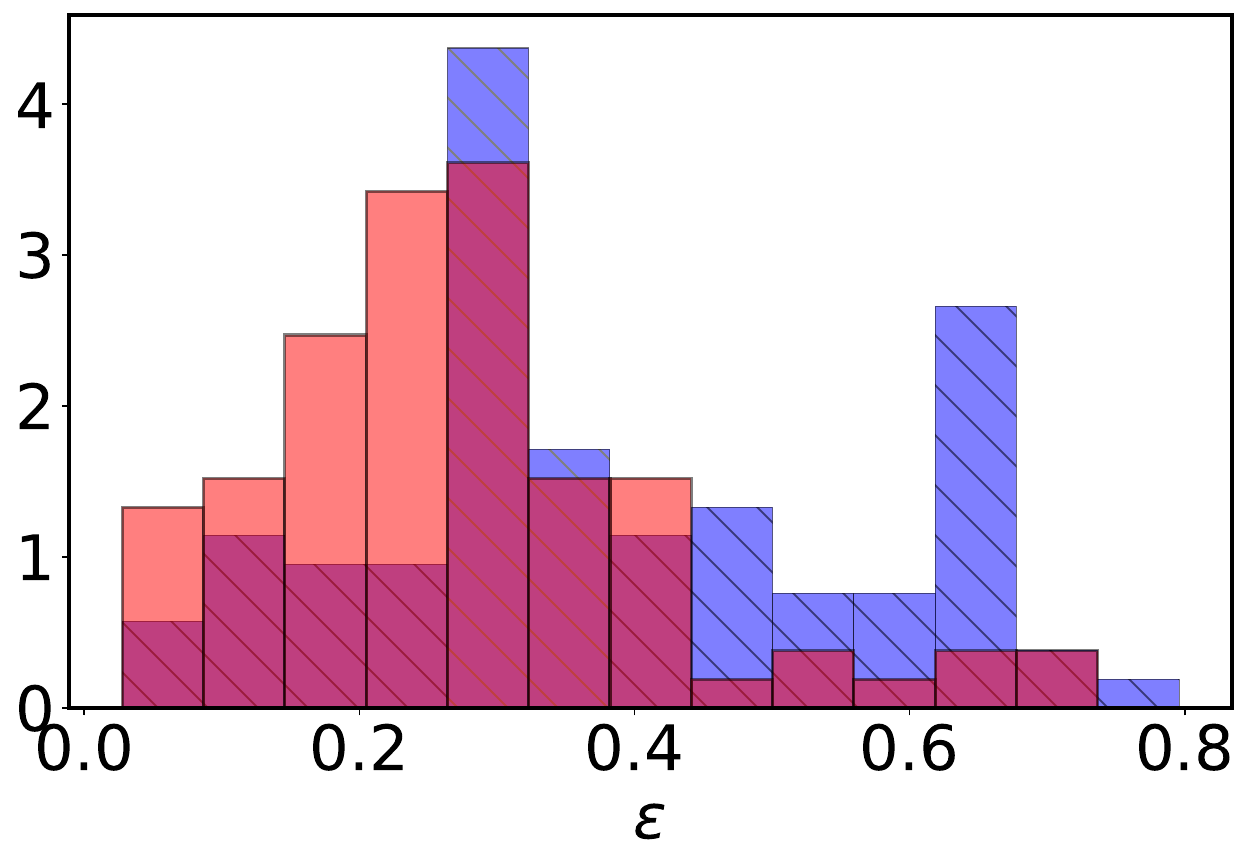} \\
\vspace{10pt} 
    \includegraphics[width=0.31\textwidth, keepaspectratio]{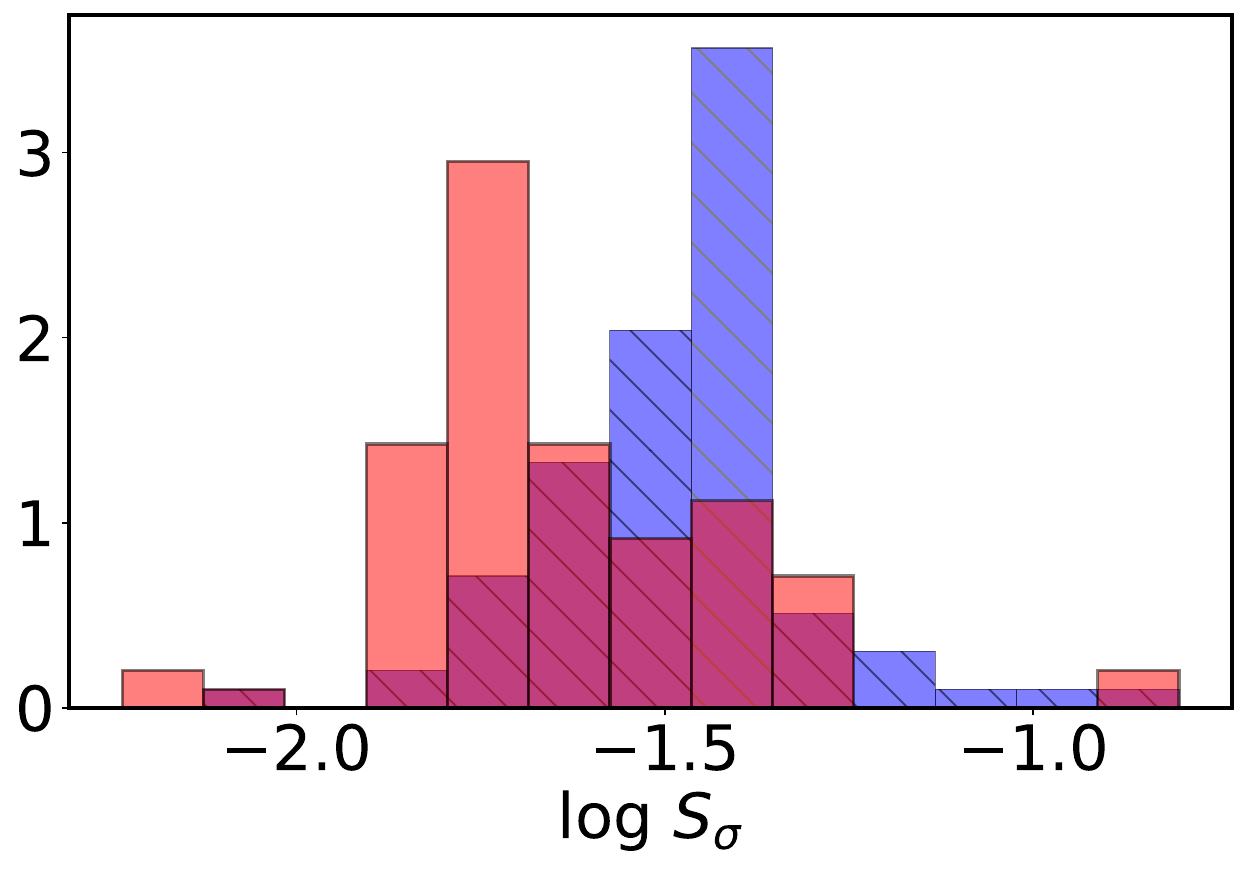} %width=0.32   % meas_log_sig_S_hist, originally: meas_log_abs_vel_skew_hist
\hspace{10pt}
    \includegraphics[width=0.31\textwidth, keepaspectratio]{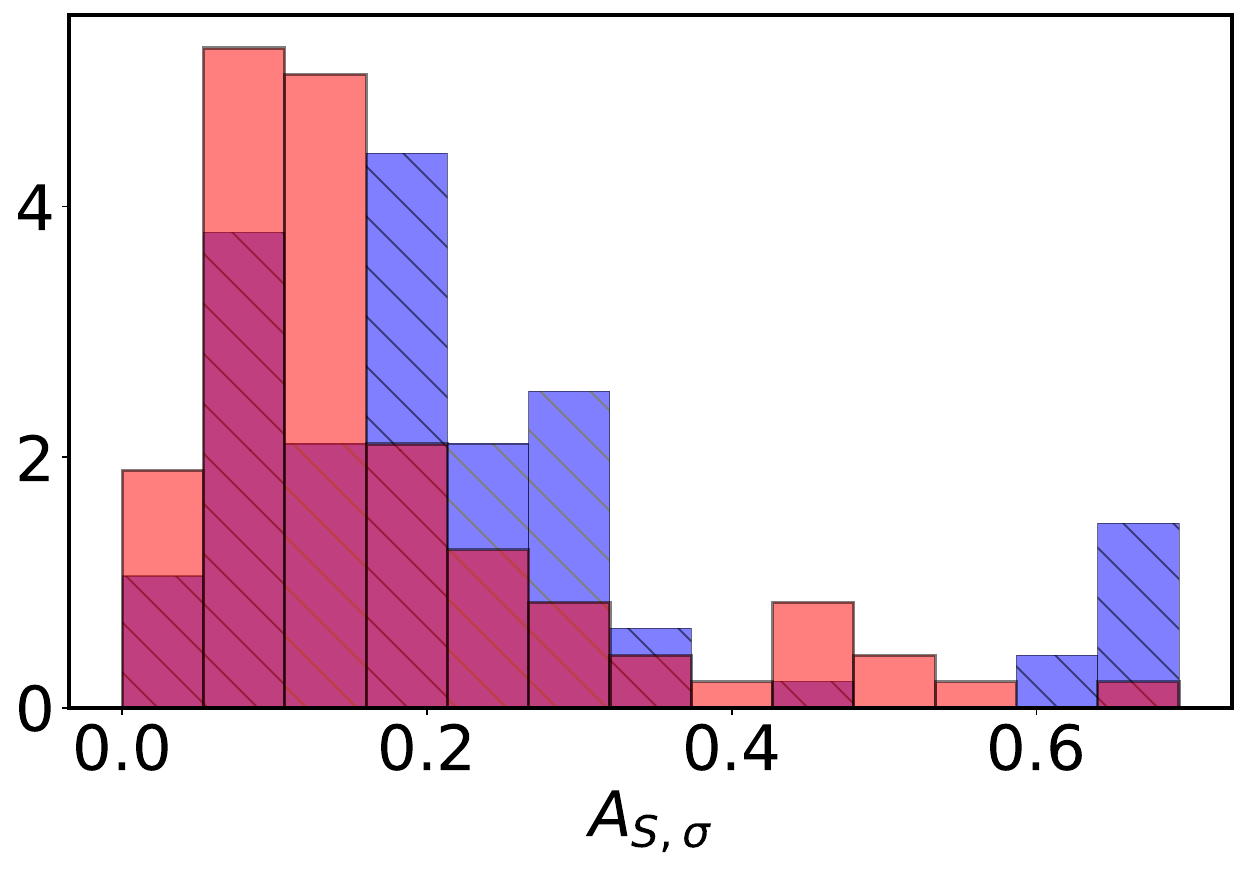} %width=0.32  % meas_sig_As_hist, originally: meas_log2_vel_kurtosis_hist
\figcaption{Normalized histograms of the kinematic parameters of the subsample of MBH mergers host galaxies (red) with high mass ratios ($\nicefrac{M_2}{M_1} \geqslant 0.5$) and high chirp masses ($M_\text{chirp} \geqslant 10^{8.2} M_\odot$), in comparison to a corresponding mass- and redshift-matched control sample (blue). We only include kinematic parameters that pass a two-sample Kolmogorov–Smirnov (K-S) test with a $>$$3\sigma$ significance. Although these histograms show significant differences between the MBH merger host galaxies and the control sample, a higher accuracy of discrimination between the two classes can be obtained by combining multiple kinematic parameters, thus motivating our use of LDA in Section~\ref{subsec:LDA}.}
\label{fig:meas_hists}
\end{figure*}

Figure~\ref{fig:kinematic_parameter_extraction} displays a visualization that summarizes the calculation of a few kinematic measurements for an example MBH merger host galaxy, including histograms of $\sigma_\star$ and $v_\mathrm{los}$ used to compute the statistical moments, the kinematic and morphological centers, and the position angles used to calculate $\Delta$PA. Figure~\ref{fig:meas_hists} displays normalized histograms of the most significant kinematic parameters for a subsample of MBH merger host galaxies with high ($\nicefrac{M_2}{M_1} \geqslant 0.5$) and high chirp masses ($M_\text{chirp} \geqslant 10^{8.2} M_\odot$), in comparison to the corresponding control sample of galaxies. This subsample of MBH merger host galaxies  with high chirp mass and mass ratio contains a total of 66 spectral datacubes, which includes datacubes generated along different lines of sight to account for inclination effects.
The histograms in Figure~\ref{fig:meas_hists} suggest that the best individual kinematic predictors for the presence of a MBH merger are $\lambda_{R_e}$, $\Delta$PA, and $\mu_{1,\sigma}$. In particular, $\lambda_{R_e}$ alone can discriminate between MBH merger host galaxies and control galaxies with $\sim$80\% accuracy. In Section~\ref{subsec:LDA}, we will show that using a linear combination of parameters results in even higher accuracies of $\gtrsim$85\%.

%%%%%%%%%%%%%%%%%%%%%%%%%%%%%%%%%%%%%%%%%%%%%%%%%%%%%
%%%%%%%%%%%%%%%%%%%%%%%%%%%%%%%%%%%%%%%%%%%%%%%%%%%%%

%%%%%%%%%%%%%%%%%%%%%%%%%%%%%%%%%%%%%%%%%%%%%%%%%%%%%
\begin{figure*}[ht]
\centering
    \includegraphics[width=0.66\textwidth, keepaspectratio]{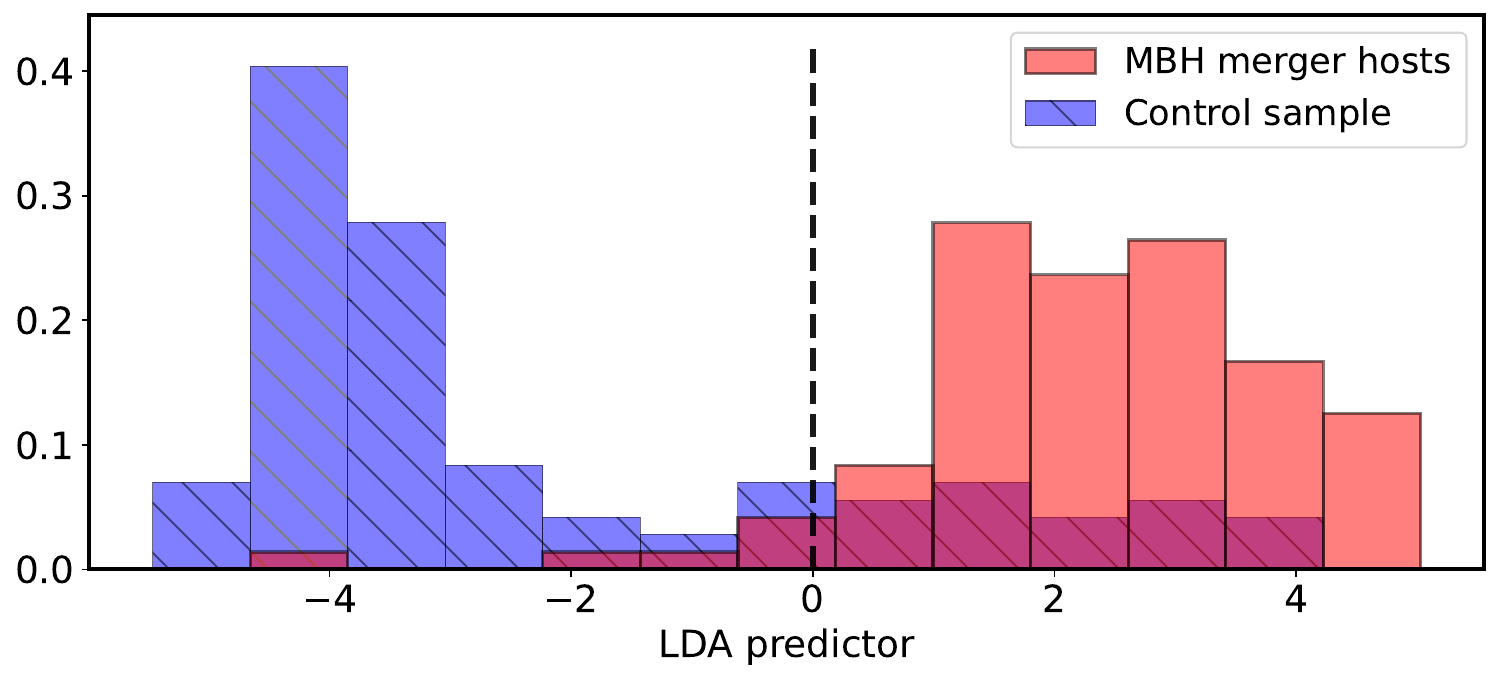} 
    \hspace{10pt}
    \includegraphics[width=0.30\textwidth, keepaspectratio]{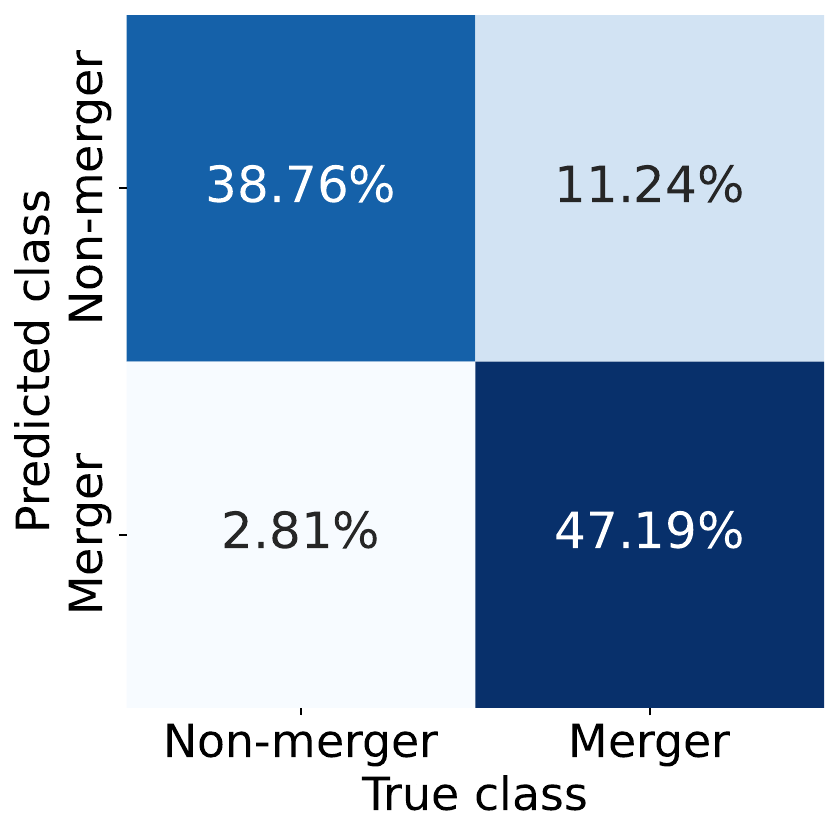}
\figcaption{ Histogram of the LDA predictor (left) for MBH merger hosts (red) and control galaxies (blue), for the same high mass and high mass ratio subsample as shown in Figure~\ref{fig:meas_hists}. 
The LDA predictor finds a linear combination of stellar kinematic parameters (Equation~\ref{eq:LD1selection}) that maximally discriminates between these two classes. The decision boundary is indicated by the dashed black line, with positive LDA predictor value indicating a MBH merger host, and negative LDA predictor value indicating a control galaxy. The confusion matrix (right) shows that for this subsample, the LDA predictor can classify galaxies with an accuracy of 86.0\% and a precision of 94.3\%.}
\label{fig:LDA_example}
\end{figure*}
%%%%%%%%%%%%%%%%%%%%%%%%%%%%%%%%%%%%%%%%%%%%%%%%%%%%%

\subsection{Linear Discriminant Analysis}\label{subsec:LDA}

We use a LDA to find a linear combination of kinematic parameters that optimally discriminates  between our MBH merger host galaxies and control sample. Our LDA approach here is the same as that used in Paper I for morphological parameters; we summarize the most salient points below, and refer the reader to Paper I for more details. The LDA finds a hyperplane in parameter space that optimally separates the two classes in the training data, and the linear equation perpendicular to this hyperplane (the `LDA predictor') provides a signed distance between each datapoint to the hyperplane that can be used for classification. The efficacy of this predictor is summarized in a confusion matrix, in which the sum of the diagonal of the confusion matrix is the accuracy (the ratio of training data predicted by the LDA correctly). We use the \texttt{sklearn} Python package \citep[][]{scikit-learn} to perform the LDA.

In addition to the stellar kinematic parameters we discussed in Section~\ref{subsec:mergerpredictors}, we also include interaction terms in the LDA, which are multiplicative combinations of morphological measurements (e.g., $\lambda_{R_e}\times\Delta \mathrm{PA}$). The inclusion of these interaction terms account for cross-correlations between parameters, thus improving the predictive ability of the LDA \citep{James_2013}, and enabling direct interpretation of the LDA coefficients in Section~\ref{subsec:coeffanalysis}.
    
To prevent any particular parameter from unphysically dominating the LDA, we first whiten (i.e., normalize) our stellar kinematic parameters. Specifically, we subtract the mean of the sample from each data point, and then divide by the standard deviation of the sample.

In training the LDA predictor, we remove stellar kinematic parameters that are uninformative in discriminating between MBH merger host galaxies and the control sample. Specifically, we only use parameters that pass a two-sample Kolmogorov–Smirnov test with a $p$-value of $\leqslant$ 0.05 ($2\sigma$) to train the LDA predictor, rejecting the hypothesis that the MBH merger host galaxy sample is drawn from the same distribution as the control sample in the histograms in Figure~\ref{fig:meas_hists}. 

To estimate uncertainties on the classification accuracies from the LDA predictor, we perform repeated stratified $k$-fold cross validation. Specifically, we divide the sample of MBH merger host galaxies and control sample into $k$ bins of equal size, where the first $k-1$ bins are used for training the LDA, and the last bin is used as a test set to test the resultant LDA. This is repeated a total of $n=10$ times, with a different set of $k$ bins. The LDA accuracy and uncertainty is thus the median and $1\sigma$ spread across all runs. In our tests, we find that $k=5$ provides a good balance between the LDA accuracy and uncertainties, and that the LDA is robust to the assumed $k$ or $n$.
    
To reduce the dimensionality of the LDA predictor and improve its interpretability, we reduce the number of parameters by training the LDA using forward stepwise selection. We first begin with a LDA model that has only a single parameter, and choose the parameter that maximizes the total LDA cross validation accuracy (i.e., minimizes the number of misclassifications). We then add additional parameters one at a time and retrain the LDA using $k$-fold cross validation iteratively, until the accuracy no longer increases. To confirm that our results are not affected by overfitting, we randomly shuffle the morphological parameters between the MBH merger host galaxies and and control samples, and verify that the LDA accuracies are consistent with $\sim$50\%. 

Figure~\ref{fig:LDA_example} shows the median-scoring trained LDA predictor for a sample of MBH merger host galaxies with high chirp masses ($M_\text{chirp} \geqslant 10^{8.2} M_\odot$) and high mass ratios ($\nicefrac{M_2}{M_1} \geqslant 0.5$), as well as its associated confusion matrix. As discussed below in Section~\ref{subsec:trends}, we specifically focus on this sample because it is the host galaxies of these MBH mergers that have the most unique stellar kinematic signatures, leading to high LDA accuracies. We find that the LDA discriminates between this MBH merger host galaxy sample and its corresponding control sample with a median accuracy of 85.7 $\pm$ 4.5\% and median precision of $93.1 \pm 6.7$\%. The linear equation of the LDA predictor in Figure~\ref{fig:LDA_example} is:
\begin{equation}\label{eq:LD1selection}
    \begin{split}
    \text{LDA predictor} = 0.51 \log\Delta\mathrm{PA} - 2.81 \lambda_{R_e} + 0.04
    % \text{LDA predictor} = 1.05\mu_{1,\sigma} + -0.82\lambda_{R_e} + -0.86\varepsilon\\ + -0.35\mu_{1,\sigma}*\varepsilon + -0.02\\
%sig mean:  		                       +1.17 +/- 0.14 **
%my spin_param:  		                   -0.85 +/- 0.15 **
%gband ellipticity_asymmetry:  		       -0.81 +/- 0.13 **
%sig mean*gband ellipticity_asymmetry:     -0.48 +/- 0.08 **
%intercept:                                -0.10 +/- 0.05
    \end{split}
\end{equation}

The cross validation also provides uncertainties on the coefficients in Equation~\ref{eq:LD1selection}. For this subsample, the 1$\sigma$ uncertainties are $\pm 0.16$ and $\pm 0.32$, for the $\log\Delta\mathrm{PA}$ and $\lambda_{R_e}$ parameters, respectively. We discuss these results in Section~\ref{sec:discussion}.\\

%\textbf{When including parameters not commonly measured in large IFU surveys such as higher moments of the line-of-sight velocity and velocity dispersion maps ($\mu_{3,v}$, $\mu_{4,v}$, $\mu_{3,\sigma}$, $\mu_{4,\sigma}$) and the morphology measurements from \texttt{StatMorph} (such as velocity dispersion smoothness and shape asymmetry $S_{\sigma}$, $A_{S, \sigma}$), we obtain a median accuracy on the high chirp mass, high mass ratio subsample of 91.4 $\pm$ 6.1\%.} The corresponding LDA predictor equation is:

%\begin{equation}\label{eq:LD1fullselection}
%    \begin{split}
%    \text{LDA predictor}_2 = 1.38\mu_{1,\sigma} %+ 0.52|\mu_{3,\sigma}| + %0.49\Delta\mathrm{PA}\\ 
%    -1.44\lambda_\mathrm{R_e} -1.41S_{\sigma} %+0.73A_{S, \sigma}
% sig mean:  		 +1.38 +/- 0.17 **
% vel abs_skew:      +0.52 +/- 0.08 **
% kin delta_PA:      +0.49 +/- 0.11 **
% my spin_param:  	 -1.44 +/- 0.20 **
% sig S:  		     -1.41 +/- 0.25 **
% sig As:  		     +0.73 +/- 0.08 **
% intercept:         +0.00 +/- 0.07 
%    \end{split}
%\end{equation}
%For this subsample, when including parameters not commonly measured in large IFU surveys (equation \ref{eq:LD1fullselection}), the mean and 1$\sigma$ uncertainties are $1.38 \pm 0.17$, $0.52 \pm 0.08$, $0.49 \pm 0.11$, $-1.44 \pm 0.20$, $-1.41 \pm 0.25$, and $0.73 \pm 0.08$, for the $\mu_{1,\sigma}$, $|\mu_{3,\sigma}|$, $\Delta\mathrm{PA}$, $\lambda_\mathrm{R_e}$, $S_{\sigma}$, and $A_{S, \sigma}$ parameters, respectively. 

%%%%%%%%%%%%%%%%%%%%%%%%%%%%%%%%%%%%%%%%%%%%%%%%%%%%%
\begin{figure*}[ht]
\centering
    \includegraphics[width=0.44\linewidth]{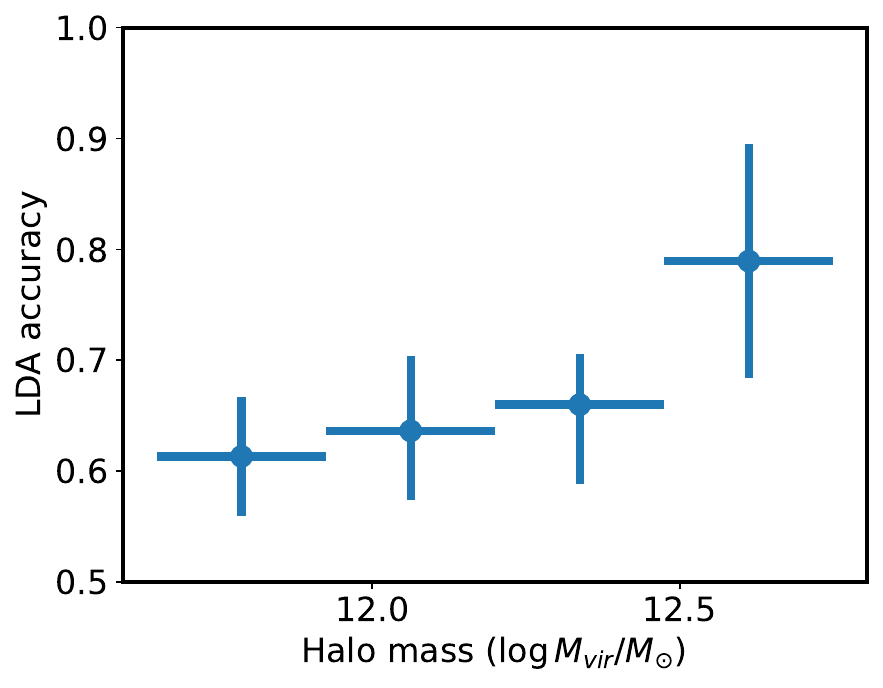}
    \hspace{10pt}
    \includegraphics[width=0.45\linewidth]{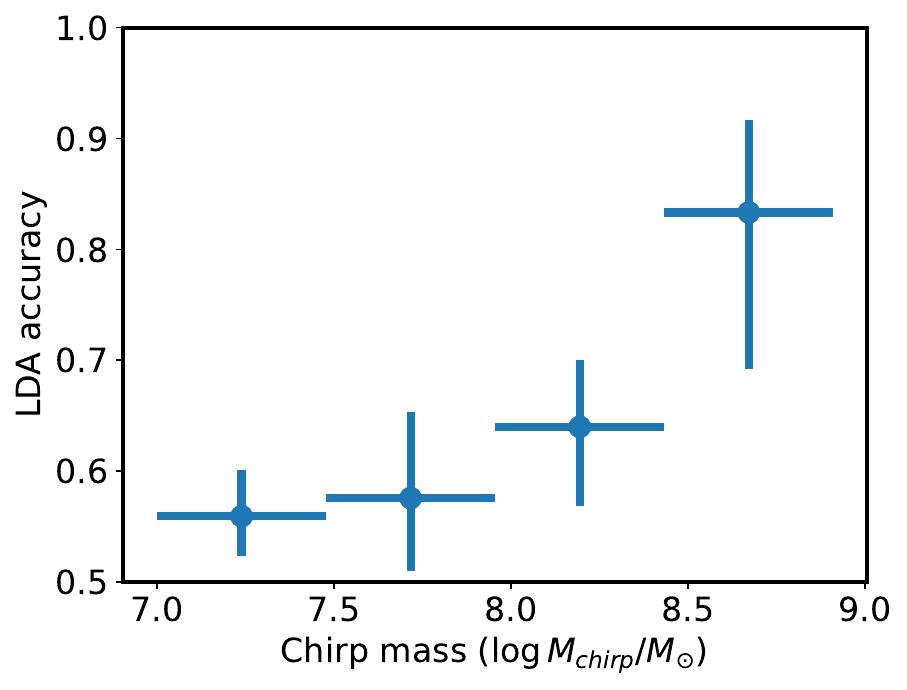} \\
    \vspace{2pt}
    \includegraphics[width=0.45\linewidth]{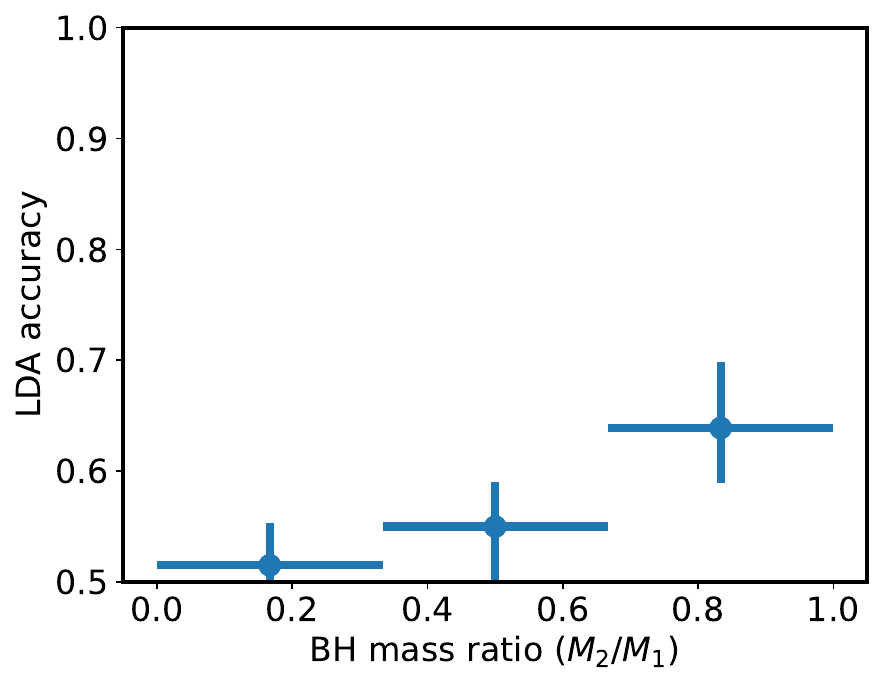}
    \hspace{10pt}
    \includegraphics[width=0.45\linewidth]{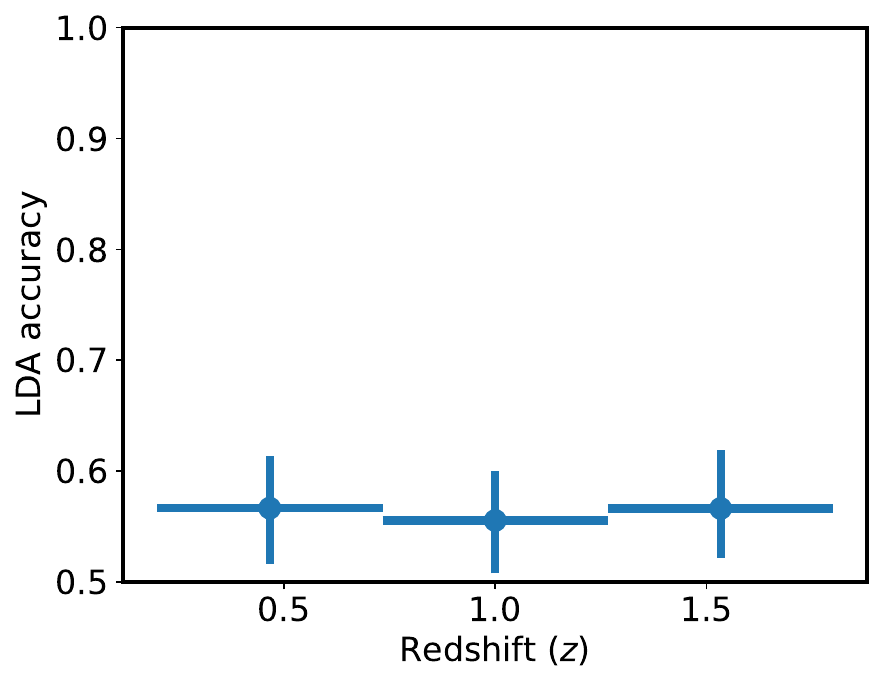}
\figcaption{The accuracy of the LDA predictor as a function of halo mass (upper left), chirp mass (upper right), MBH mass ratio (lower left), and redshift (lower right) of the MBH merger in the Romulus25 simulation. For visual clarity, the halo mass, chirp mass, and redshift plot only contain MBH mergers with high mass ratio ($\nicefrac{M_2}{M_1} \geqslant 0.5$), while the mass ratio plot only contains MBH merger with high chirp masses of $M_\text{chirp} \geqslant 10^{7.5} M_\odot$. The error bars are the 1$\sigma$ spread from cross validation. The accuracy of our stellar kinematics based approach increases as a function of halo mass, chirp mass, and mass ratio, and does not have a significant trend with redshift.}
\label{fig:ldascore_vs_cuts}
\end{figure*}
%%%%%%%%%%%%%%%%%%%%%%%%%%%%%%%%%%%%%%%%%%%%%%%%%%%%% 

%%%%%%%%%%%%%%%%%%%%%%%%%%%%%%%%%%%%%%%%%%%%%%%%%%%%%
\section{Results and Discussion} \label{sec:discussion}
 
\subsection{LDA Coefficient Analysis}\label{subsec:coeffanalysis}
The most distinct kinematic parameters identified by the LDA in Equation~\ref{eq:LD1selection} are the spin parameter $\lambda_{R_e}$, and difference between kinematic and photometric position angles $\Delta \mathrm{PA}$. Specifically, the histograms in Figure~\ref{fig:meas_hists} show that MBH merger host galaxies have systematically lower $\lambda_{R_e}$ (slower spin) and higher $\Delta \mathrm{PA}$ (stronger kinematic misalignments), relative to the control sample. Formally, a two-sample K-S test shows that the distributions of $\lambda_\mathrm{R_e}$ and $\Delta \mathrm{PA}$ for MBH merger host galaxies is statistically distinct from the control sample in Figure~\ref{fig:meas_hists} at a $\gtrsim$$10\sigma$ and 6.1$\sigma$ level, respectively. 

%%%%%% Note that the new KS-Test values for the values shown in figure 3 are 6.6, 3.5, 5.3, 7.1, 6.0, 3.0, 5.1 sigma level for the sigma mean, sigma absolute skew, delta PA, spin parameter, sigma smoothness, sigma shape asymmetry, and gband ellipticity, respectively. %%%%%%%

Figure~\ref{fig:meas_hists} also shows that MBH merger host galaxies have systematically higher mean velocity dispersion $\mu_{1, \sigma}$, at high significance of 6.4$\sigma$ in a two-sample K-S test. However, $\mu_{1, \sigma}$ was not chosen by the LDA for inclusion in Equation~\ref{eq:LD1selection} during training, because it is correlated with $\lambda_\mathrm{R_e}$, and thus its inclusion in Equation~\ref{eq:LD1selection} would not add additional discriminating power. This latent correlation is observed in IFU surveys of nearby galaxies, which show that the most massive galaxies have lower $\lambda_\mathrm{R_e}$ \citep{Emsellem_2011, Falcon-Barroso_2019}.

\subsection{Trends in Discriminating Power}\label{subsec:trends}
We investigate how the accuracy of the LDA predictor in Equation~\ref{eq:LD1selection} depends on properties such as the halo virial mass, chirp mass, mass ratio, and redshift. To do this, we divide both our MBH merger host galaxies and corresponding control samples into subsamples (e.g., in bins of chirp mass). For each subsample, we retrain the LDA predictor to calculate its accuracy. The resulting trends are shown in Figure~\ref{fig:ldascore_vs_cuts}. 

Figure~\ref{fig:ldascore_vs_cuts} shows that the LDA accuracy increases with halo mass, chirp mass, and mass ratio, and is relatively insensitive to redshift. These trends from stellar kinematics are similar to those for morphological signatures of these same MBH merger host galaxies that we previously found Paper I. We discuss the origin of these trends in Section~\ref{subsec:interpretation} below.

\subsection{Dependence of Results on Time-Delay}\label{subsec:timescales}
\begin{figure}[tb]
\begin{center}
    \includegraphics[width=0.45\textwidth, keepaspectratio]{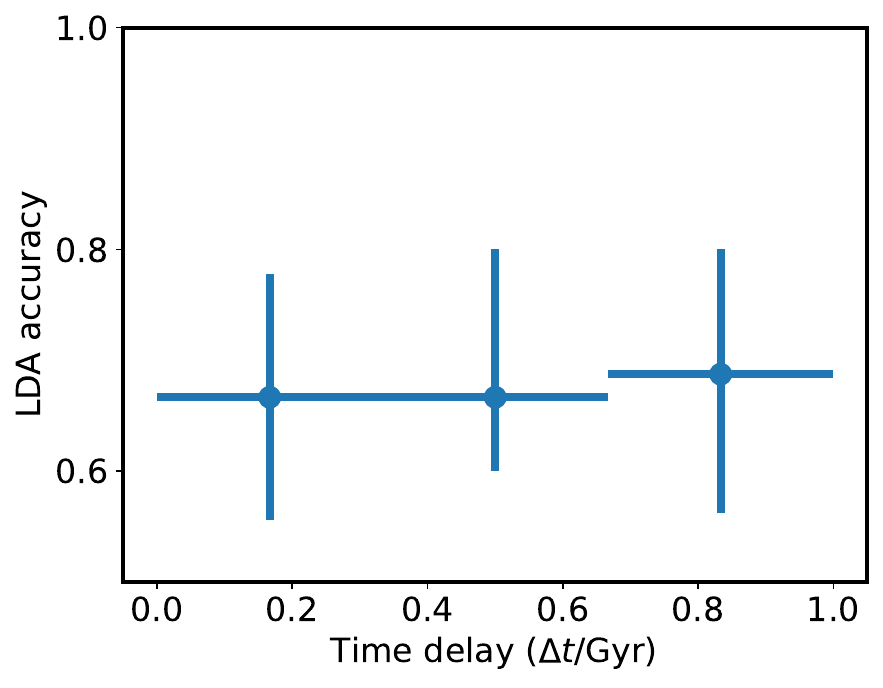}
\end{center}
\figcaption{The accuracy of the LDA predictor as a function of time-delay since numerical merger in the simulation, for a high mass ratio ($\nicefrac{M_2}{M_1} \geqslant 0.5$) and high chirp mass ($M_{chirp} > 10^{8} M_\odot$) MBH merger subsample and corresponding control sample. The LDA accuracy remains high for at least $\sim$1~Gyr, suggesting that the unique stellar kinematic signatures of MBH merger host galaxies are likely to be permanent. Thus, the accuracy of the stellar kinematic approach to identifying MBH merger host galaxies will likely be high at the time of gravitational wave detection.}
\label{fig:ldascore_vs_deltat}
\end{figure}

We test whether the accuracy of the LDA predictor changes as a function of time-delay between the numerical merger and the physical merger of MBHs. For each binary in our simulation, the time-delay between the numerical merger in the simulation (at binary separations of $\sim$700~pc) and the physical merger (when a gravitational wave chirp would be emitted) is highly uncertain, and can range from 0.1 to 10 Gyrs \citep[e.g.,][]{Volonteri_2020, Li_2022}. Instead of carefully estimating this time-delay for each merger in our samples, we remain agnostic to the detailed physics below the resolution limit of the simulation, and instead test whether the LDA accuracy changes up to 1~Gyr after the numerical merger. Figure~\ref{fig:ldascore_vs_deltat} shows the LDA accuracy as a function of time-delay, for the same high mass ($\gtrsim$10$^{8.2}$~$M_\odot$) and high mass ratio ($\gtrsim$0.5) subsample from Figure~\ref{fig:LDA_example}. The accuracy of our trained LDA predictor remains high for at least $\sim$1~Gyr after numerical merger (see Section~\ref{subsec:interpretation}).

%%%%%%%%%%%%%%%%%%%%%%%%%%%%%%%%%%%%%%%%%%%%%%%%%%%%%

%%%%%%%%%%%%%%%%%%%%%%%%%%%%%%%%%%%%%%%%%%%%%%%%%%%%%

%%%%%%%%%%%%%%%%%%%%%%%%%%%%%%%%%%%%%%%%%%%%%%%%%%%%%
\begin{figure*}[tb]
\begin{center}
    \includegraphics[width=0.48\textwidth, keepaspectratio]{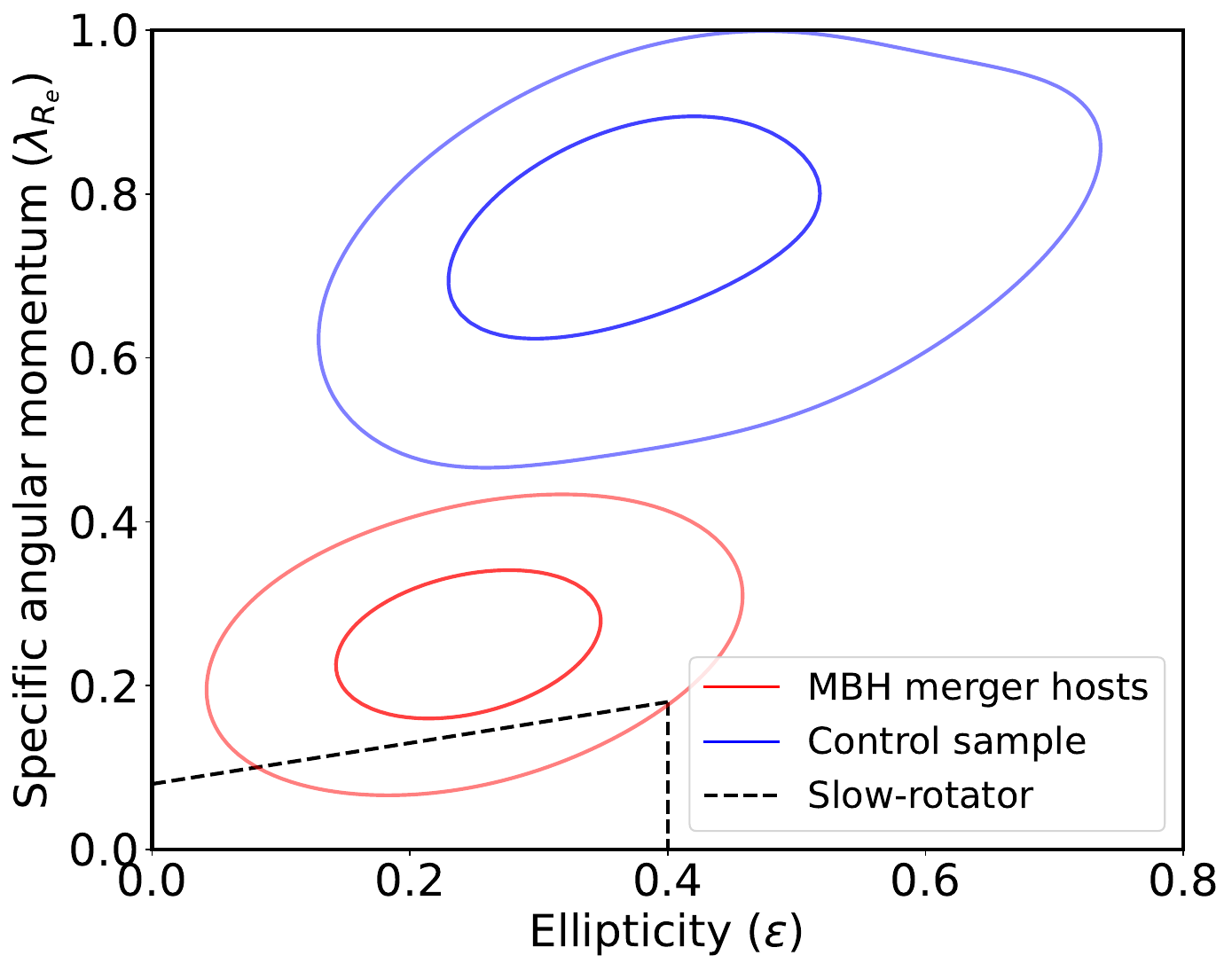}
    \includegraphics[width=0.49\textwidth, keepaspectratio]{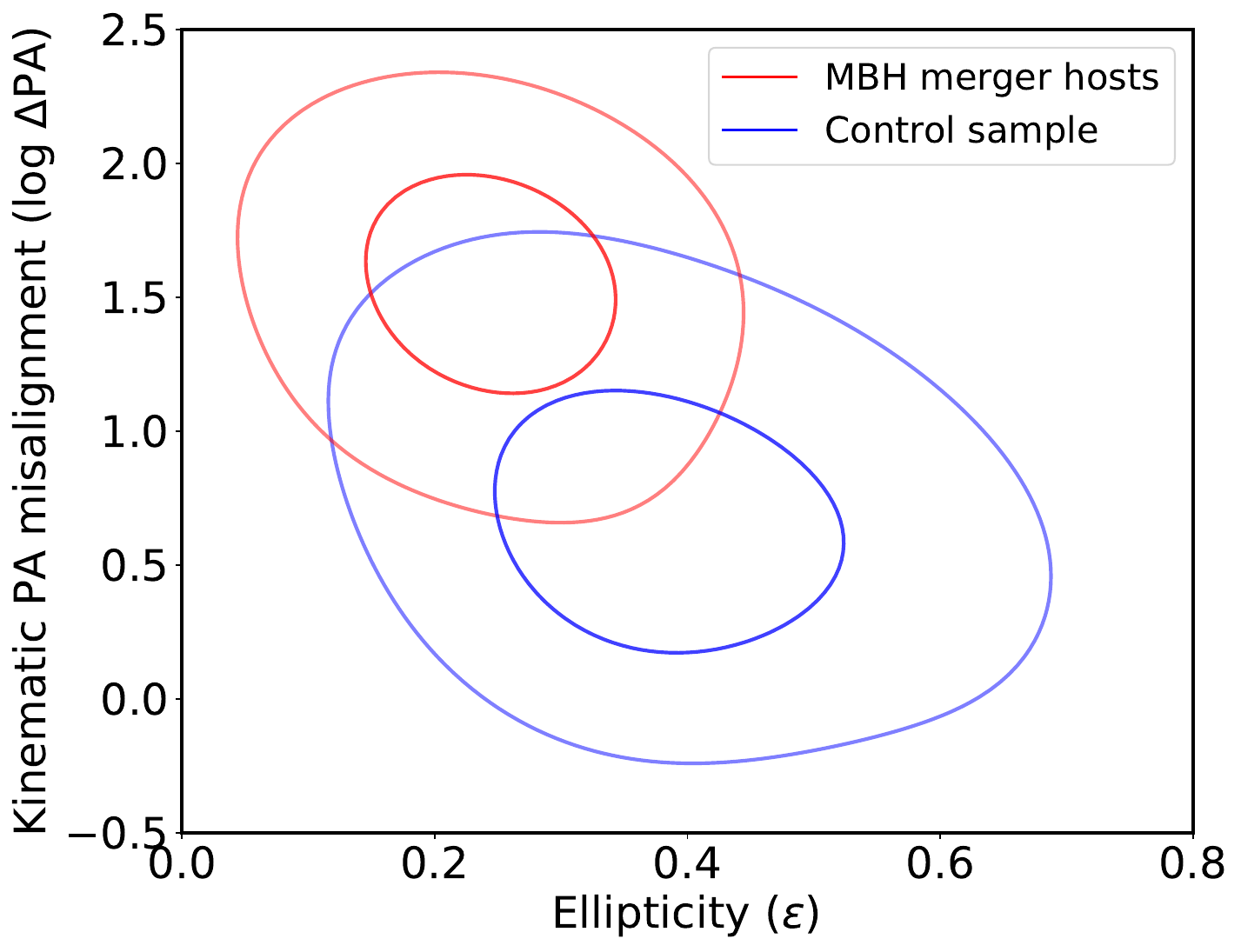}
\end{center}
\figcaption{ {\it Right panel:} A comparison of the spin parameter $\lambda_{R_e}$ and morphological ellipticity $\varepsilon$, for MBH merger host galaxies and the control sample. MBH merger host galaxies have systematically slower rotation (smaller $\lambda_{R_e}$) and are more round (smaller $\varepsilon$). 
{\it Left panel:} A comparison of the difference between kinematic and photometric position angles $\Delta \mathrm{PA}$ and morphological ellipticity $\varepsilon$, for the same MBH merger host galaxies and control sample. MBH merger host galaxies are systematically more kinematically misaligned (larger $\Delta \mathrm{PA}$).
Galaxies with all of these kinematic properties are thought to have undergone major mergers, which causes kinematic misalignments and slower rotation. For visual clarity, both panels show the subsample of high mass ratio MBH merger host galaxies($\nicefrac{M_2}{M_1} \geqslant 0.5$) with high chirp masses ($M_\text{chirp} \geqslant 10^{8.2} M_\odot$) (red contours), with a corresponding mass- and redshift-matched control sample (blue contours). The contours enclose 15.9\% and 50\% of each sample. The dashed black line is the slow rotator discriminant from \citet{Cappellari_2016}, in which galaxies with small values of $\lambda_{R_e}$ and $\varepsilon$ below the line are classified as slow rotator galaxies. 
}
\label{fig:spin_deltaPA_contours}
\end{figure*}
 %%%%%%%%%%%%%%%%%%%%%%%%%%%%%%%%%%%%%%%%%%%%%%%%%%%%%
 
\subsection{Interpretation and Comparison to Results from Morphology in Imaging}\label{subsec:interpretation}

Our stellar kinematic results support the view that the most distinctive characteristics of MBH binary and merger host galaxies are permanent features, rather than transient disturbances stemming from the preceding galaxy mergers. Previously in Paper I, we used the same MBH merger host galaxy and control galaxy samples from Romulus25 that we use here to study their morphologies in synthetic broadband images produced from dust radiative transfer. In that study, we extracted morphological parameters from these synthetic images and trained an LDA predictor, similar to our use of kinematic parameters to train an LDA predictor in Section~\ref{sec:analysis} above. Paper I showed that the accuracy of the resultant morphology-based LDA predictor does not decrease as a function of time-delay, which suggests that the unique morphological signatures of MBH merger host galaxies are permanent, instead of transient features such as short-lived morphological disturbances stemming from the preceding galaxy merger. This is consistent with our results shown in Figure~\ref{fig:ldascore_vs_deltat}, which demonstrates that the accuracy of our LDA predictor trained on the stellar kinematic parameters of these galaxies also does not decrease over time for at least 1~Gyr after the numerical merger. Thus, both morphology and stellar kinematics indicate that MBH merger host galaxies have unique and long-lived characteristics.

The stellar kinematics of our simulated MBH merger host galaxies also suggest that their most distinctive characteristics are permanent kinematic properties associated with major mergers of massive galaxies, consistent with our morphological results from Paper I. The two most informative stellar kinematic parameters in our LDA predictor for MBH merger host galaxies are a systematically slower spin $\lambda_{R_e}$, and a larger difference between the kinematic and photometric position angles $\Delta \mathrm{PA}$, as discussed in Section~\ref{subsec:coeffanalysis} above and shown in Figure~\ref{fig:meas_hists}. Galaxy IFU surveys have revealed galaxies with both slow spin \citep[small $\lambda_{R_e}$;][]{Emsellem_2011, Graham_2018, Ene_2020} and strong misalignment between their kinematic and photometric position angles \citep[large $\Delta \mathrm{PA}$;][]{Krajnovic_2011, Cortese_2016, Ene_2018}. These properties are more clearly shown for our simulated galaxies in Figure~\ref{fig:spin_deltaPA_contours}, which compares our MBH merger host galaxies to the control sample in the $\lambda_{R_e}$ -- $\varepsilon$ and $\Delta \mathrm{PA}$ -- $\varepsilon$ planes, as commonly plotted for IFU observations. Numerical simulations show that these galaxy properties result from major galaxy mergers, which cause irregular rotation, and also decrease the net rotation of galaxies \citep[e.g.,][]{Jesseit_2009, Bois_2011, Khochfar_2011, Naab_2014, Penoyre_2017, Lagos_2018}. Furthermore, we also find in Figure~\ref{fig:meas_hists} that MBH merger hosts also have systematically higher mean stellar velocity dispersion $\sigma_*$, although this parameter was not selected by the LDA for inclusion in Equation~\ref{eq:LD1selection} because it is correlated with $\lambda_\mathrm{R_e}$, and thus its predictive information is redundant. Higher $\sigma_*$ is indicative of more prominent bulges, which are also a natural consequence of major galaxy mergers. Overall, our interpretation that the unique  stellar kinematic signatures of MBH merger host galaxies stem from major mergers of massive galaxies is highly consistent with our previous results in Paper I, which showed that the unique morphological signatures of MBH merger host galaxies is the presence of a strong classical bulge, built through major galaxy mergers.

Our interpretation also naturally explains the trends in the LDA accuracy that we observe as a function of chirp mass and mass ratio. A key result from this study is that the stellar kinematics of MBH merger host galaxies are only distinct for MBH binaries and mergers with high chirp masses and high mass ratios. These trends are clearly observed in Figure~\ref{fig:ldascore_vs_cuts}, and naturally arise from a scenario in which the unique kinematic signatures owe to major mergers of massive galaxies. Numerical simulations demonstrate that galaxies with slow spin and large kinematic misalignments arise from major mergers of massive gas-poor galaxies \citep[e.g.,][]{Jesseit_2009, Bois_2011, Naab_2014}. Since it is specifically these galaxy mergers that produce MBH binaries and mergers with high chirp mass and mass ratio, this naturally causes the LDA accuracy to increase with chirp mass and mass ratio, as observed in Figure~\ref{fig:ldascore_vs_cuts}.
Since chirp mass is correlated with halo mass, it is thus also expected that the LDA accuracy is high for the most massive halos in Figure~\ref{fig:ldascore_vs_cuts}. Although the LDA accuracy from our previous results from host galaxy morphology decreased mildly with redshift, our LDA accuracy from stellar kinematics in Figure~\ref{fig:ldascore_vs_cuts} does not appear to display strong trends with redshift. This may owe to the fact that bulges are more difficult to resolve and detect in the morphology of higher-redshift galaxies, while their stellar kinematic signatures in IFU spectroscopy remain relatively distinct.

A consistent picture thus emerges from our study of the unique stellar kinematics of MBH merger host galaxies, in broad agreement with our previous results from morphology in Paper I, as well as other studies. As massive galaxies undergo major mergers, they form prominent classical bulges that can be observed in their morphologies in broadband imaging (e.g., through the bulge discriminant statistic $F(Gini, M_{20})$ described in Paper I) and directly through larger mean stellar velocity dispersions $\sigma_*$ in IFU spectra. The history of major mergers in these galaxies also causes their stellar kinematics to display misalignments  (large $\Delta \mathrm{PA}$) and slower rotation (lower $\lambda_{R_e}$). The MBHs in these massive merging galaxies will then go on to form binaries with high chirp masses and mass ratios, and eventually merge due to gravitational wave losses. It is thus specifically in MBH binaries and mergers with high chirp masses and mass ratios that both the morphological and stellar kinematic signatures of the host galaxies are most unique. These signatures can thus be used to identify the host galaxies of MBH binaries and mergers detected in gravitational waves with high accuracies. Lastly, our results are also broadly consistent with other studies that use cosmological simulations to investigate the host galaxies of MBH binaries detectable by PTA experiments (which preferentially have high mass ratios and high chirp masses). In particular, \citet{Saeedzadeh_2024} show that PTA-detectable MBH binaries in Romulus25 are massive early-type galaxies with low star formation rates residing at the centers of galaxy groups, while \citet{Cella_2024} show that PTA-detectable MBH binaries in the Illustris simulations have redder colors and higher metallicities. These properties are all expected from major mergers of massive galaxies, similar to the stellar kinematic and morphological signatures that we identify.

\subsection{Practical Considerations for Observations}\label{subsec:observations}

We caution that our trained LDA predictor may not be directly applicable to observations from IFU instruments with very different specifications than assumed here. In producing our synthetic spectral datacubes, we assumed specifications such as medium spectral resolution of $R = 4000$,  pixel scales of $0\farcs1$/pixel, and Gaussian $0\farcs1$ FWHM PSFs. The LDA predictor in Equation~\ref{eq:LD1selection} can thus be directly applied to observed spectral datacubes obtained by similar instruments. However, our results may no longer hold true when applied to spectral datacubes from telescopes and instruments with vastly different specifications. We defer a more thorough test of these effects to future work. 

Despite our caution above, we suggest that in practice, our results may actually be more broadly applicable even to more modest instruments and observing conditions. The fact that both morphological and stellar kinematic host galaxy signatures are only distinct for MBH binaries and mergers with high chirp masses and mass ratios suggest that these approaches will be most useful for gravitational wave detections of individual MBH binaries by PTA experiments. These MBH binaries will be in relatively nearby galaxies, at significantly lower redshifts than our sample of simulated host galaxies from Romulus25. Thus, current IFU instruments with more modest capabilities may already be able to achieve results comparable to our findings, by resolving physical scales in each galaxy that are similar to our assumptions. For example, at the median redshift ($z \sim 1$) of simulated galaxies in our sample, the assumed $0\farcs1$ pixels resolve physical scales of approximately 0.8~kpc, while this same physical scale can be resolved with $2\farcs$0 pixels for a nearby galaxy at 90~Mpc. This suggests that our trained LDA predictor could be directly applied to IFU spectral datacubes of nearby galaxies that have significantly worse angular resolutions, provided that they probe similar physical scales in the host galaxy targets. Furthermore, archival IFU spectra of a significant percent of these nearby galaxies already exist from large IFU surveys such as SAURON \citep{Bacon_2001}, ATLAS3D \citep{Cappellari_2011}, CALIFA \citep{Sanchez_2012}, MASSIVE \citep{Ma_2014}, SAMI \citep{Bryant_2015}, and MANGA \citep{Bundy_2015}. Thus, efforts to identify MBH merger host galaxies in gravitational wave localization volumes using our stellar kinematics approach can leverage existing archival data,  minimizing the need to obtain expensive new IFU spectroscopy.

The main disadvantage of using a stellar kinematics approach to identifying MBH binary and merger host galaxies is that it will only be useful if the number of candidate host galaxies lying in typical gravitational wave error volumes is small.
For a gravitational wave error volume, selection cuts based on redshift and stellar mass (assuming empirical scaling relations) must be used first to limit the number of candidate galaxies, before additional approaches such as galaxy stellar kinematics or morphology can be applied. Despite the high accuracies of $\gtrsim$85\% we find using our stellar kinematics approach, it will only be useful if there are a small number ($\lesssim$10) of candidate galaxies remaining in the gravitational wave error volume. Such a scenario is possible for PTA detections of individual binaries in optimistic scenarios \citep{Petrov_2024}, but in practice a combination of several approaches will be required to identify the host galaxy.

\subsection{Caveats}\label{subsec:caveats}

Although we tested the accuracy of our LDA predictor in Figure~\ref{fig:ldascore_vs_deltat} up to 1~Gyr after each numerical merger, the time-delay to physical merger can be up to several Gyrs. For each MBH merger, the time-delay between the numerical merger and physical merger is highly uncertain, and is dependent on the poorly-understood dynamical friction and stellar scattering below the resolution limit of the simulation. More detailed calculations have suggested that this time-delay can range between 0.1--10~Gyr \citep{Volonteri_2020, Li_2022}. In Section~\ref{subsec:timescales}, we remain agnostic to the exact time-delay of each MBH merger, and instead show in Figure~\ref{fig:ldascore_vs_deltat} that the LDA accuracy does not decrease for at least $\sim$1~Gyr. This choice was primarily driven by the computational costs of performing radiative transfer simulations. We also place no additional selection criteria on the control sample, to avoid artificially limiting the diversity in the merger histories of our galaxies, thus ensuring that our results reflect the realistic population of galaxies formed in a cosmological context. The constant accuracy of our LDA predictor over $\sim$1~Gyr after numerical merger is consistent with (and an expected consequence of) our result that the most unique characteristics of MBH merger host galaxies are slower spin, stronger kinematic misalignment, and larger mean velocity dispersions, since these are relatively permanent features of galaxies. Nevertheless, future work could explicitly test these results beyond 1~Gyr, to understand whether these unique signatures eventually fade.

%%%%%%%%%%%%%%%%%%%%%%%%%%%%%%%%%%%%%%%%%%%%%%%%%%%%%
\section{Conclusions} \label{sec:conclusions}

We demonstrated that the host galaxies of MBH binaries and mergers detected in gravitational wave experiments have unique stellar kinematics in IFU spectra. Using a sample of simulated galaxies hosting MBH mergers selected from a cosmological simulation, we performed stellar population synthesis and dust radiative transfer simulations to produce optical 3D spectral datacubes that are similar to data from IFU spectrographs. We performed full-spectrum fitting on these spectral datacubes to generate kinematic maps of $\sigma_\star$ and $v_\mathrm{los}$ for each host galaxy. We then extracted a variety of stellar kinematic parameters from these maps, and compared their properties to a mass- and redshift-matched control sample of simulated galaxies that do not host MBH mergers. Specifically, we trained a LDA predictor on the stellar kinematic parameters, to produce a linear equation that optimally distinguishes MBH merger host galaxies from the control sample. Our main findings are:

\begin{enumerate}
    \item The stellar kinematic properties of MBH merger host galaxies are unique, suggesting that IFU spectra can be used to identify the host galaxies of MBH binaries and mergers detected in gravitational waves. The accuracy of this approach increases with chirp mass and mass ratio; for mergers with particularly high chirp masses ($\gtrsim$10$^{8.2}$ $M_\odot$) and high mass ratios ($\gtrsim$0.5), the accuracies reach $\gtrsim$85\%. These accuracies are slightly higher than approaches using galaxy morphology in broadband imaging, with similar trends. However, our results also suggest that in practice, both stellar kinematic and morphology-based approaches to identifying host galaxy counterparts will only be useful for the most massive MBH binaries and mergers, such as those that will be detected in nearby galaxies by PTA experiments.
    
    \item The most unique stellar kinematic properties of MBH merger host galaxies are systematically lower specific angular momentum $\lambda_{R_e}$, larger difference between kinematic and photometric position angles $\Delta \mathrm{PA}$, and higher stellar velocity dispersions $\sigma_*$. This implies that that MBH merger host galaxies have systematically slower rotation, more irregular stellar kinematics, and stronger bulges. Observationally, these properties are commonly associated with massive early-type galaxies that have undergone major mergers, which would naturally host MBH binaries and mergers. Our results are thus consistent with a scenario in which MBH binaries and mergers with high masses and mass ratios form in major mergers of massive galaxies, which have unique morphologies and stellar kinematics as probed by broadband imaging and IFU spectroscopy.
    
\end{enumerate}

The high accuracies of $\gtrsim$85\% achieved by our stellar kinematics approach suggests that IFU spectroscopy can play a role in telescope follow-up of MBH binaries and mergers detected in gravitational waves. These $\gtrsim$85\% accuracies are slightly higher than the $\gtrsim$80\% accuracies we find in Paper I using host galaxy morphology in broadband imaging. However, we emphasize that we find these high accuracies specifically for MBH binaries and mergers with high chirp masses and mass ratios, and thus our stellar kinematics approach may only be relevant for PTA detections of massive MBH binaries. Nevertheless, these PTA detections are limited to nearby massive galaxies, where IFU spectroscopy of galaxies in the localization region will be most feasible, and a plethora of archival IFU spectra already exists.

Our results suggest that prospects are bright for multi-messenger science with MBH binaries and mergers that will be detected in low-frequency gravitational waves. When a MBH binary or merger with high chirp mass and mass ratio is detected, initial redshift and stellar mass selection cuts will reduce the candidate galaxies in the gravitational wave error volume by orders of magnitude. If only a small number ($\lesssim$10) of candidates remain, additional broadband imaging and IFU spectroscopy may uniquely identify the host galaxy based on morphology and stellar kinematics, respectively. Crucially, these approaches will be effective even if there are no other electromagnetic signatures, such as if there is no active accretion onto the MBH binary or merger, or if there is heavy dust obscuration. By combining insights from a variety of host galaxy identification techniques, the promise of the diverse array of science goals based on multi-messenger observations of MBH binaries and mergers can be realized.

%%%%%%%%%%%%%%%%%%%%%%%%%%%%%%%%%%%%%%%%%%%%%%%%%%%%%
\begin{acknowledgments}

We thank Thomas R.\ Quinn, Jessie Runnoe, Stephen Taylor, Kelly Holley-Bockelmann, and Maria Charisi for insightful discussions.

This work made extensive use of the \href{https://docs.alliancecan.ca/wiki/Cedar}{Cedar} cluster of \href{https://alliancecan.ca/}{Digital Research Alliance of Canada} at Simon Fraser University.

Romulus25 is part of the Blue Waters sustained-petascale computing project, which is supported by the National Science Foundation (awards OCI-0725070 and ACI-1238993) and the state of Illinois. Blue Waters is a joint effort of the University of Illinois at Urbana-Champaign and its National Center for Supercomputing Applications.

 J.B.\ acknowledges support from the NSERC Summer Undergraduate Research Award program. J.J.R.\ and D.H.\ acknowledge support from the Canada Research Chairs program, the NSERC Discovery Grant program, and the FRQNT Nouveaux Chercheurs Grant program. J.J.R.\ acknowledges funding from the Canada Foundation for Innovation, and the Qu\'{e}bec Ministère de l’\'{E}conomie et de l’Innovation.
 
\end{acknowledgments}

%%% === SOFTWARE === %%%
\software{
\href{https://github.com/SKIRT/SKIRT9}{\texttt{SKIRT}}: \cite{Camps_2015};
\href{https://github.com/pynbody/pynbody}{\texttt{Pynbody}}: \cite{Pynbody};
\href{https://github.com/pynbody/tangos}{\texttt{Tangos}}: \cite{Tangos};
\href{https://github.com/cconroy20/fsps}{\texttt{FSPS}}: \cite{Conroy_Gunn_2010};
\href{https://pypi.org/project/ppxf/}{\texttt{ppXF}}: \cite{Cappellari_23};
\href{https://github.com/vrodgom/statmorph}{\texttt{StatMorph}}: \cite{Rodriguez-Gomez_2019};
\href{https://scikit-learn.org}{\texttt{sklearn}}: \cite{scikit-learn};
\href{https://github.com/N-BodyShop/changa}{\texttt{ChaNGa}}: \cite{Menon_2015};
\href{http://popia.ft.uam.es/AHF/}{\texttt{Amiga Halo Finder}}: \cite{Knollmann_2009};
\href{https://www.astropy.org/}{\texttt{astropy}}: \cite{astropy18}
}
%%%%%%%%%%%%%%%%%%%%%%%%%%%%%%%%%%%%%%%%%%%%%%%%%%%%%

\bibliography{paper}{}
\bibliographystyle{aasjournal}

\end{document}